\documentclass[preprint,3p,twocolumn,11pt]{elsarticle}

\usepackage{amssymb}
\usepackage{booktabs}
\usepackage{dcolumn}
\usepackage{multirow}
\usepackage{adjustbox}
\usepackage{subfig}
\usepackage{graphicx}
\usepackage{natbib} 
\usepackage[obeyspaces,hyphens]{url}
\usepackage{booktabs}
\usepackage{lscape}
\usepackage{nicefrac}
\usepackage{xcolor}
\usepackage[countmax]{subfloat}

\biboptions{sort&compress}
\newcommand{\be}{\begin{equation}}
\newcommand{\ee}{\end{equation}}
\newcommand{\bse}{\begin{subequations}}
\newcommand{\ese}{\end{subequations}}
\newcommand{\beq}{\begin{eqnarray}}
\newcommand{\eeq}{\end{eqnarray}}

\usepackage{amsmath}
\usepackage{txfonts}
\usepackage{epsfig}
\usepackage{graphicx}
\usepackage{empheq}
\journal{}

\begin{document}

\begin{frontmatter}

\title{World personal income distribution evolution measured by
purchasing power parity exchange rates}

\author[1]{J.D.A.\ Islas-Garc\'{\i}a\fnref{fn1}}
\address[1]{Posgrado Ciencias F\'{\i}sicas, Universidad Nacional
     Aut\'{o}noma de M\'{e}xico, Mexico City, Mexico}
\ead{j\_da\_ig@ciencias.unam.mx}

\author[2]{M.\ del Castillo-Mussot\fnref{fn2}}
\address[2]{Instituto de F\'{\i}sica, Universidad Nacional Aut\'{o}noma
     de M\'{e}xico, Mexico City, Mexico}
\ead{mussot@fisica.unam.mx}

\author[3]{Marcelo B.\ Ribeiro\fnref{fn3}}
\address[3]{Instituto de F\'{\i}sica, Universidade Federal do Rio de
     Janeiro, Rio de Janeiro, Brazil}
\ead{mbr@if.ufrj.br}
\fntext[fn1]{Orcid 0000-0002-2440-0936}
\fntext[fn2]{Orcid 0000-0001-5061-2601}
\fntext[fn3]{Orcid 0000-0002-6919-2624}

\begin{abstract}
The evolution of global income distribution from 1988 to 2018 is analyzed
using purchasing power parity exchange rates and well-established statistical
distributions. This research proposes the use of two separate distributions
to more accurately represent the overall data, rather than relying on a
single distribution. The global income distribution was fitted to log-normal
and gamma functions, which are standard tools in econophysics. Despite
limitations in data completeness during the early years, the available
information covered the vast majority of the world's population. Probability
density function (PDF) curves enabled the identification of key peaks in the
distribution, while complementary cumulative distribution function (CCDF)
curves highlighted general trends in inequality. Initially, the global income
distribution exhibited a bimodal pattern; however, the growth of middle
classes in highly populated countries such as China and India has driven the
transition to a unimodal distribution in recent years. While single-function
fits with gamma or log-normal distributions provided reasonable accuracy,
the bimodal approach constructed as a sum of log-normal distributions yielded
near-perfect fits.
\end{abstract}

\begin{keyword}
econophysics \sep
income distribution \sep
gamma function \sep
log-normal distribution 
\end{keyword}
\end{frontmatter}


\section{Introduction}

The increasing interconnectedness of the global economy over the past four
decades has sparked intense interest in understanding how income is distributed
across the world's population. This interest stems from the growing realization
that globalization has often been accompanied by rising income and wealth
inequalities \cite{oxfam1,oxfam2}. Inasmuch as both income and wealth
distributions, and their respective inequalities, go to the heart of any
society's viewpoints on issues regarding egalitarianism and social opportunity,
the relationship between globalization processes and its impact on the income
inequality has attracted a lot of recent interest, becoming in fact an
important research topic among economists and econophysicists
\cite{piketty,incomedistro,ribeiro2020income}. In particular, the
characterization of income distributions yield critical information for
determining richness, the gap between rich and poor and societies' well
being rates at any {\it gross domestic product} (GDP) level
\cite{elephantpaper2}.

Research aimed at determining the overall behavior of income
distributions were initially focused on some countries and regions
and within limited time period intervals. More recently such an
approach, although much more detailed than earlier studies, still
focuses on particular countries and regions, which renders such
studies basically fragmented if one considers the worldwide income
distribution \cite{ribeiro2020income} and, therefore, do not present
a general scenario of its global situation. This is a clearly
desirable goal in order to advance our understanding of the
income distribution dynamic evolution at the world scale.

The aim of establishing a global income distribution must, however,
rely upon the combination of as many national household surveys as
possible because there is no global household survey of individual
incomes. To include all countries is not an easy task, as discussed
by Milanovic \cite{milanovic2002true,milanovic2016greatest}, and
more recently by Anand and Segal \cite{anand2015global}, because
most of the first works focusing on the world income distribution
are studies of international inequality in the sense that they
calculated what would be inequality in the world if it were
populated by representative individuals from all countries, that
is, by people having the mean income of their countries.

More accurate representations of the world income distributions were
constructed afterwards from assembling income distributions of
countries obtained by using income surveys or tax data. As mentioned
by Milanovic \cite{milanovic2016greatest}, global inequality is a
relatively recent topic because in order to calculate it one needs
to have data on national income distributions for most of the
countries in the world, or at least for most of the populous
countries. Only from the early to mid 1980s that such data became
available for China, the Soviet Union and its constituent republics,
as well as large parts of Africa. Nevertheless, the problem of data
homogeneity, which  ensures that variables are defined the same way
as much as possible, has been a difficult one in this area since its
inception.

The world income, or expenditure distribution, for 1988 and 1993
was calculated in Ref.\ \cite{milanovic2002true} for individuals
based entirely on household surveys from 91 countries adjusted for
differences in \textit{purchasing power parity} (PPP) between
countries covering about 84\% of world population and 93\% of
world GDP. In similar works \cite{bourguignon2002size,
berry1983changes} income shares for a number of countries were
approximated using income shares of ``similar'' countries. More
recently, making use of household income data from more than 130
countries the evolution of the global income distribution between 2008
and 2013 after the financial crisis was analyzed
\cite{milanovic2022after}. For a comprehensive review of many
recent aspects of the global distribution of income, including
conceptual and methodological issues, inequality and global poverty
, we refer to the works of Milanovic
\cite{milanovic2016greatest,milanovic2006global,
milanovic2011worlds,milanovic2012global,milanovic2016income,
milanovic2023great} and Anand and Segal \cite{anand2015global,
anand2008we,anand2010debates,anand2017global,segal2022inequality}.

This paper aims at fitting the global income distribution data over
several years and studying it evolution. Here we follow the tradition
of economists, physicists and mathematicians who have sought to
characterize the distribution of income in countries by a mixture of
known statistical distributions \cite{ribeiro2020income}. Our approach
here is to try to characterize the changes in time of the individual
income distribution in the world as a whole by means of known statistical
distributions with the smallest possible number of parameters.

The ultimate aim of studies on income and wealth distribution must be
to reveal the inner dynamics of both quantities by expressing them in
terms of time evolving differential equations \cite{ribeiro2020income}.
So, the ultimate aim must be to identify the mechanisms at work so that
some further theoretical work clarifies and enhances our understanding
of what we observe \cite{piketty}. However, income distribution is a
subject that was unfortunately very much neglected by mainstream academic
economics for a very long time \cite{atkinson97}, and whose revival basically
happened on the onset of the 21st century \cite{ribeiro2020income}, so the
present research level of this subject still very much remains in the
stage of data collecting and analyzing in order to see which basic
conclusions can be reached from the data in order to try to point out
possible future theoretical endeavors. This is particularly true of
global income distribution, which means that the present study is very
much focused on this initial research stage.

The plan of the paper is as follows.  \S 2 is devoted to briefly
review some models of wealth and income distributions used by
economists, econophysicists and other scientists that will be
employed in the present approach. The exponential-like distributions
such as the gamma and log-normal distributions. Since our data are at
household-level (micro) data, then in \S 3 we briefly present the
limitations of the databases we employed in terms of their sources,
standardization, drawbacks and advantages, together with the
convenience of PPP to compare overall consumption and income between
nations. In \S 4 we present our results of world income distribution
between 1988 and 2018 measured by PPP in US dollars. \S 5 is devoted
to our conclusions.

\section{Modeling income and wealth distributions used in econophysics}

It is fair to state that at present there is a consensus among most,
if not all, researchers devoted to the income distribution problem
that the richest stratum of a country income distribution, that is,
its upper end segment, is well represented by a power law as Vilfredo
Pareto argued over a century ago \cite{pareto1964cours}.
However, the distributive characterization of the not so rich
still remains an open problem. Different authors proposed different
fitting functions to characterize the income distribution of the
vast majority of populations, but until the turn of the 21st century
little more has been done than trying different functional fits in
relatively limited number of countries or group of countries
\cite{incomedistro,ribeiro2020income}.

Among the early attempts at different functional fits one should
recall the work of Robert Gibrat, who in 1931 had already indicated
that the Pareto law is only valid for the high income range, whereas
for the small and middle income ranges he suggested the log-normal
probability density as a better descriptor. He also proposed a law
of proportionate effect, which states that a small change in a
quantity is independent of the quantity itself \cite{gibrat1931inegalits}.

An important, and much more recent, work in this respect was the
analysis of the income distribution data of the USA as studied by
Silva and Yakovenko \cite{yakovenko2005two}. It revealed the
coexistence of two social classes as far as functional fitting is
concerned: the large majority of the population is characterized
by a quasi-exponential distribution, and the very small upper income
segment exhibits the Pareto power-law distribution with characteristic
fat tails. They argued that there is a similar-to-physics energy
conservation law such that in the income distribution problem
translates itself as conservation of money. This means that the
middle and lower income populations of the USA are described by an
exponential function whose interpretation is of being a Boltzmann-Gibbs
distribution which entails such a conservative money quantity.

Silva and Yakovenko also considered currency transactions as being
equivalent to elastic molecular collisions in a gas particle, where
in principle all the conserved energy, in this case money, would be
transferred from one particle, or agent, to another in an one-to-one
interaction, or transaction, without money loss \cite{silva2004temporal}.
The income distribution data of Mexico \cite{soriano2017non}, the
European Union \cite{jagielski2013modelling}, and more than 60
countries \cite{tao2019exponential} also present similar two-classes
structure.

Chakrabarti and collaborators \cite{incomedistro,ribeiro2020income,joseph2022}
extended this kinetic collision model to include savings, which then
better reflects real economic transactions, yielding, in the case of
a constant saving fraction for all agents, a stationary distribution
very similar to the gamma function.

In general, the bulk of the lower distribution stratum of both income
and wealth can be fitted by exponential, log-normal and gamma
distributions. Nevertheless, contrary
to the lower regions which remain basically unchanged for both
income and wealth, apart from the different functional fits, the
Pareto tail slope exhibits changes in time, a behavior that could
be possibly explained by the complex processes of creation and
destruction of money through investments, credit, financial derivatives,
big stock market crisis, etc, features which are much more clearly
related to the Pareto tail because these processes are basically from
where the rich people extract their income and wealth \cite{ribeiro2020income}. 

\subsection{Distribution functions}

The \textit{cumulative distribution function} (CDF) is defined as
follows,
\begin{equation}
F(m) = \int_{0}^{m} P(m') \,d m', 
\label{CDFEq}
\end{equation}
where $P$ signifies the \textit{probability distribution function} (PDF),
also known as \textit{probability density}. In the present context $m$
represents monetary value. The complement of Eq.\ (\ref{CDFEq}) defines
the \textit{complementary cumulative distribution function} (CCDF), which
may be written as below \cite{ribeiro2020income}, 
\begin{equation}
\bar{F}(m) = \int_{m}^{\infty} P(m') \,d m' = 1 - F(m).
\label{CCDFEq} 
\end{equation}
This is a very useful quantity to study income distribution because it
provides valuable insights on the data it represents by offering better
visualization of tail behavior, which in turn highlights rare events
given by extreme values in the dataset, that is, far from the mean. In
addition, several CCDFs plots provide helpful comparisons on how the
tails behave, allowing the assessment of the heavier or lighter ones.
Discussing the income distribution by means of the CCDF provides a
meaningful way to comprehend the probability of values greater than
or equal to a given threshold \(m\), facilitating a deeper understanding
of the data's behavior and tail characteristics.\footnote{Eq.\
(\ref{CCDFEq}) has several applications when it is integrated in time,
specially in, but not limited to, medicine and engineering. In the
medical literature the CCDF is known as \textit{survival function}
\cite{survive}, whereas in the engineering literature
it is referred as \textit{reliability function} \cite{reli}.
In these two applications the CCDF gives the probability that a patient
survives, or a device remains reliable, past a certain time.}

\subsection{Log-normal distribution}

This is basically a normal function whose independent variable $x$
scales as $\ln x$. That is, the log-normal distribution is a normal
one of the logarithm of $x$ \cite{ribeiro2020income,kleiber2003statistical}.
So, the probability density of the normal function scaled that way may be
written as,
\begin{equation}
N(\ln x)=\frac{1}{\sigma\sqrt{2\pi}}\exp{\left[-\frac{(\ln x -\mu)^2}
{2\sigma^2}\right]},
\label{lognormal}
\end{equation}
where the parameters $\mu$ and $\sigma$ are respectively the mean value of
the logarithmic variable and its variance, that is, $\mu=\langle \ln x
\rangle$ and $\sigma = \langle (\ln x -\mu)^2 \rangle$. A change of variables
produces,
\begin{equation}
N(\ln x)\,d (\ln x)=\left[ \frac{N(\ln x)}{x} \right] \,d x.
\label{lognormal2}
\end{equation}
For equal probabilities under the normal and log-normal densities,
incremental areas should also be equal, that is, $N(\ln x)\,d (\ln x)
= N_l(x)\,d x$. This means that the probability density of the
log-normal distribution is given by,
\begin{equation}
N_l(x)=\frac{N(\ln x)}{x}=\frac{1}{x\sigma\sqrt{2\pi}}
\exp{\left[-\frac{(\ln x -\mu)^2}{2\sigma^2}\right]}.
\label{lognorm}
\end{equation}

\subsection{Gamma distribution}

The income distribution of the less than rich can also be reasonably well
fitted by the gamma distribution \cite{ribeiro2020income,
ferrero2004statistical,ferrero2005monomodal}]. This is a three-parameter
function whose probability density reads as, 
\begin{equation}
f(x)=\left[\frac{A}{\Gamma(n)m^n}\right]x^{(n-1)}{\mathrm{e}}^{-({x}/{m})},
\label{GammaEq}
\end{equation}
where $A$, $n$, and ${1}/{m}$ are, respectively, the normalizing, shape,
and rate parameters.

\subsection{Distributions constructed as sums}

During the process of data fitting we found useful to summing up
two log-normal or gamma distributions with different parameter
values. Hence, the bi-gamma PDF is written as,
\begin{multline}
f(x)=A_1x^{(n_1-1)}\left[\frac{{\mathrm{e}}^{-({x}/{m_1})}}
{\Gamma(n_1)m_1^{(n_1-1)}}\right] \\ +A_2x^{(n_2-1)}\left[
\frac{{\mathrm{e}}^{-({x}/{m_2})}} {\Gamma(n_2)m_2^{(n_2-1)}}\right],
\label{bigama}
\end{multline}
whereas the bi-log-normal PDF yields,
\begin{multline}
f(x)=\frac{A_1}{x\sigma_1} \exp{\left[-\frac{(\ln x-\mu_1)^2}
{2{(\sigma_1)}^2}\right]} \\ +\frac{A_2}{x\sigma_2}
\exp{\left[-\frac{(\ln x-\mu_2)^2}{2{(\sigma_2)}^2}\right]}.
\label{bilognormal}
\end{multline}

Statistical distributions such as the log-normal and gamma distributions
are well suited for modeling income and wealth because of their ability
to capture the natural variability of economic systems. The log-normal
distribution effectively represents middle- and low-income segments,
where incomes grow multiplicatively through investment or other economic
processes. In contrast, the gamma distribution is particularly well suited
for modeling lower-income populations because of its exponential decay but
it is vanishing at the origin. The introduction of a bimodal fit, where
two distributions are combined, allows for a more accurate representation
of the data by taking into account the coexistence of different income
groups within global populations.

\section{Database}

The GDP is a widely used monetary measure of the market value of all
final goods and services produced in a period of time, often annually.
There are two ways to measure GDP: nominally or via PPP. The first
way, nominal or market value GDP, or GDP at exchange rate, occurs when
the GDP of countries in their corresponding currencies are converted
into a single currency, like, for example, into the \textit{United
States dollar} (USD). The second measure is GDP at PPP (GDP-PPP), when
a ``basket of goods'' comprising a wide range of goods and services is
priced equally in different countries and territories and by taking
into account exchange rate.

In what follows, for brevity reasons, when we write countries, it
is understood that we refer to both countries and territories. The
so-called ``international dollar'' would buy in a given country a
comparable amount of goods and services a USD would buy in the US
according to PPP data. Although estimating the PPP across countries
is not an easy task, it is accepted that PPP measures are generally
regarded as better and more stable way than market values to compare
overall consumption and income among nations. The GDP-PPP of
developing countries is in general, higher than their nominal GDP,
so the per capita income gap between rich and poor countries is
reduced under PPP values.

The empirical data used here were obtained from two sets of data: Lakner
and Milanovic \cite{lakner2015global} and Roser \cite{Roser2023-et}. As
it will be shown below, our main fitting results were derived from the
former's database, whereas the latter's one was employed to subtract
the income distributions of China and India in order to show how
important these countries populations are to fill the ``global
middle-class'' valley as time passes.

All aforementioned data do not include the entire world population
because it has a 60k USDs upper cutoff limit that considered the
inflation adjusted year 2011. In addition, the income values were
measured in each country according to PPP in USD. Milanovic's data
\cite{lakner2015global} were measured in 2011 PPP USD and Roser's
data \cite{Roser2023-et} in 2005 PPP USD. Even so, despite this
limitation the available data included the vast majority of the
world's population. The results of our analysis will demonstrate how
simple statistical models can effectively capture the dynamic evolution
of income distributions over time. The following section delves into
these results, highlighting key patterns and through our fitting methods.

\section{Results}

Data fitting using the distribution functions discussed above were
carried out with all available data. The results are presented below
grouped by the respective function used to fit the data. In all figures
the term ``semilogx'' at the top indicates that there is a logarithmic
scale at the x-axis. 

\subsection{Gamma and bi-gamma fits}

\begin{figure}[ht]
    \centering
    \includegraphics[width=\linewidth]{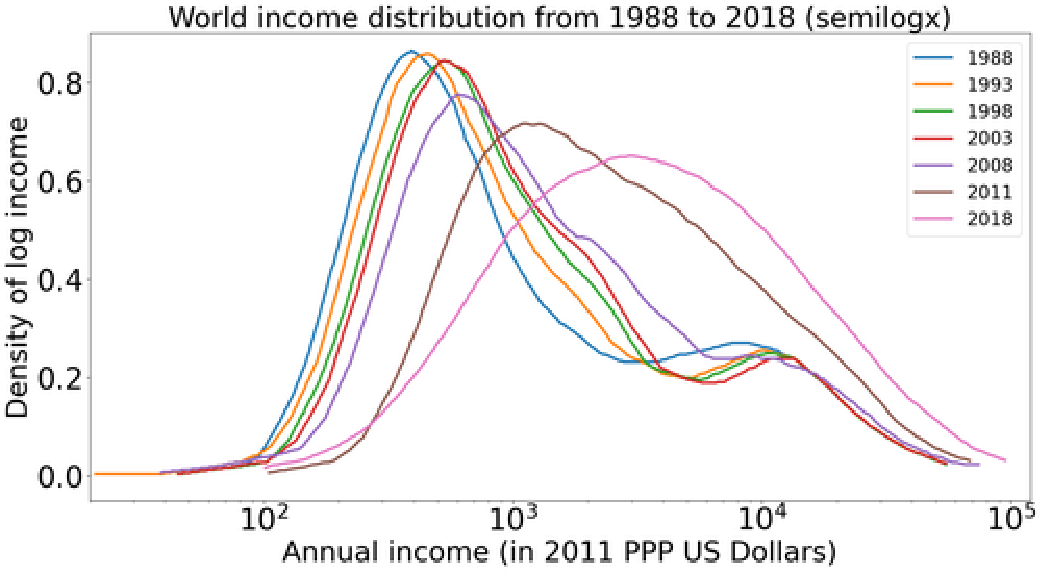}
    \caption{Milanovic income distribution from 1988 to 2018.}\label{fig:BDat}
\end{figure}

Fig.\ \ref{fig:BDat} shows the world income PDF in 2011 PPP USD from
1988 to 2018 obtained by Milanovic. Two results can be clearly noticed
from the plots as time passes: the distribution shifts to the right
and the valley tends to disappear.

\begin{figure}[ht]
    \centering
    \includegraphics[width=\linewidth]{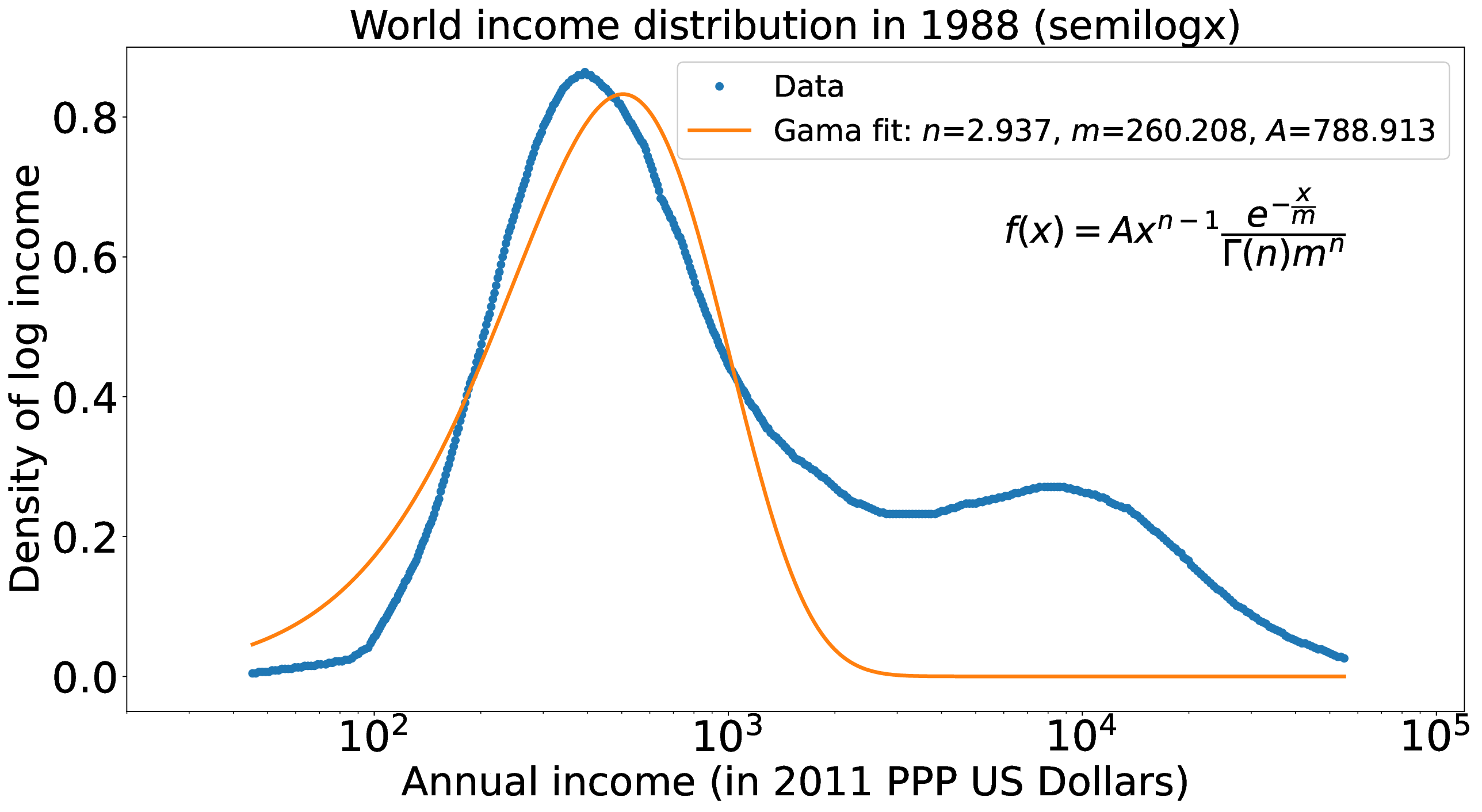}
    \caption{Gamma fit for Milanovic income distribution for year 1988, $R^2=0.69068$}\label{fig:Gama88}
\end{figure}

\begin{figure}[ht]
    \centering
    \includegraphics[width=\linewidth]{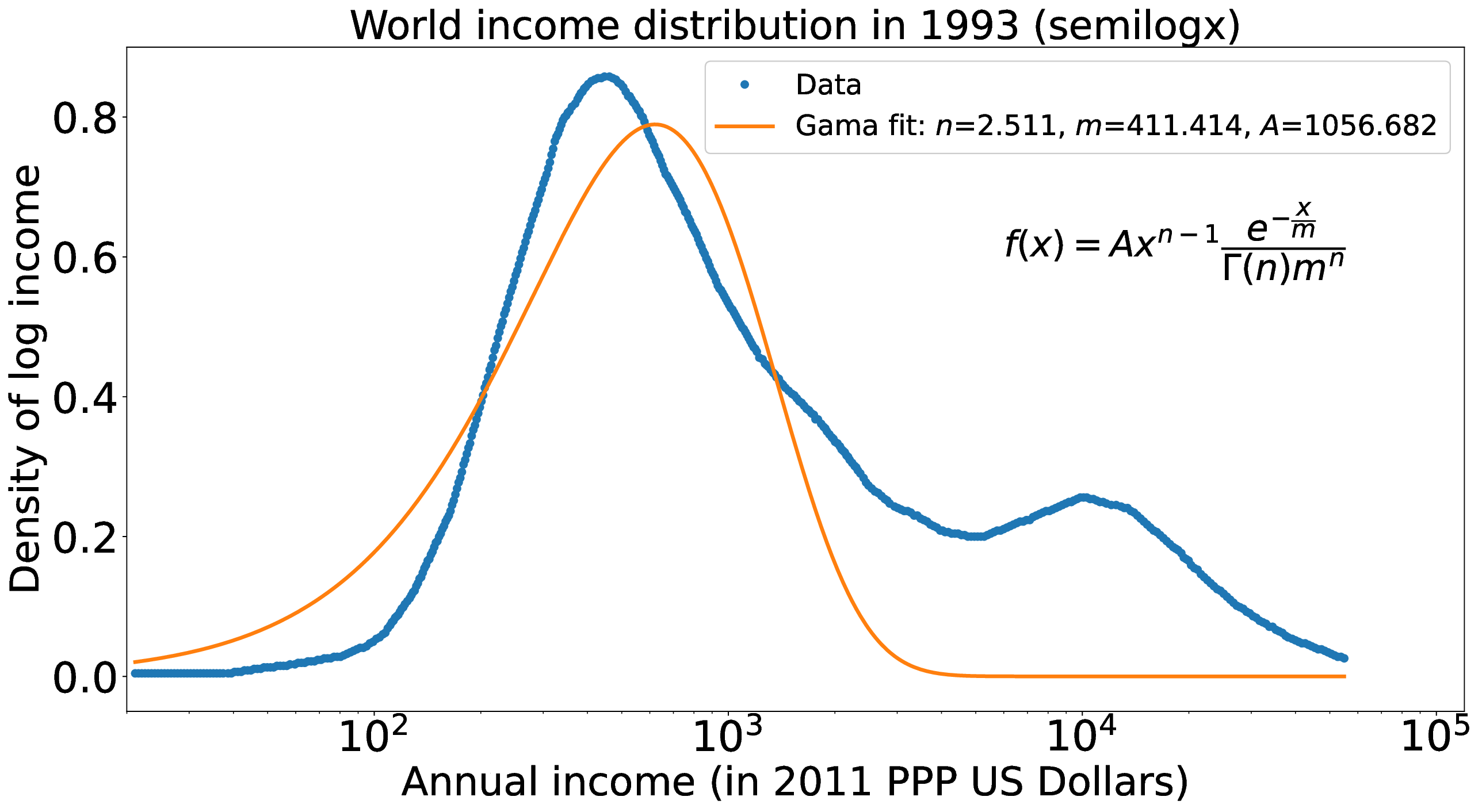}
    \caption{Gamma fit for year 1993, $R^2=0.73948$}\label{fig:Gama93}
\end{figure}

\begin{figure}[ht]
    \centering
    \includegraphics[width=\linewidth]{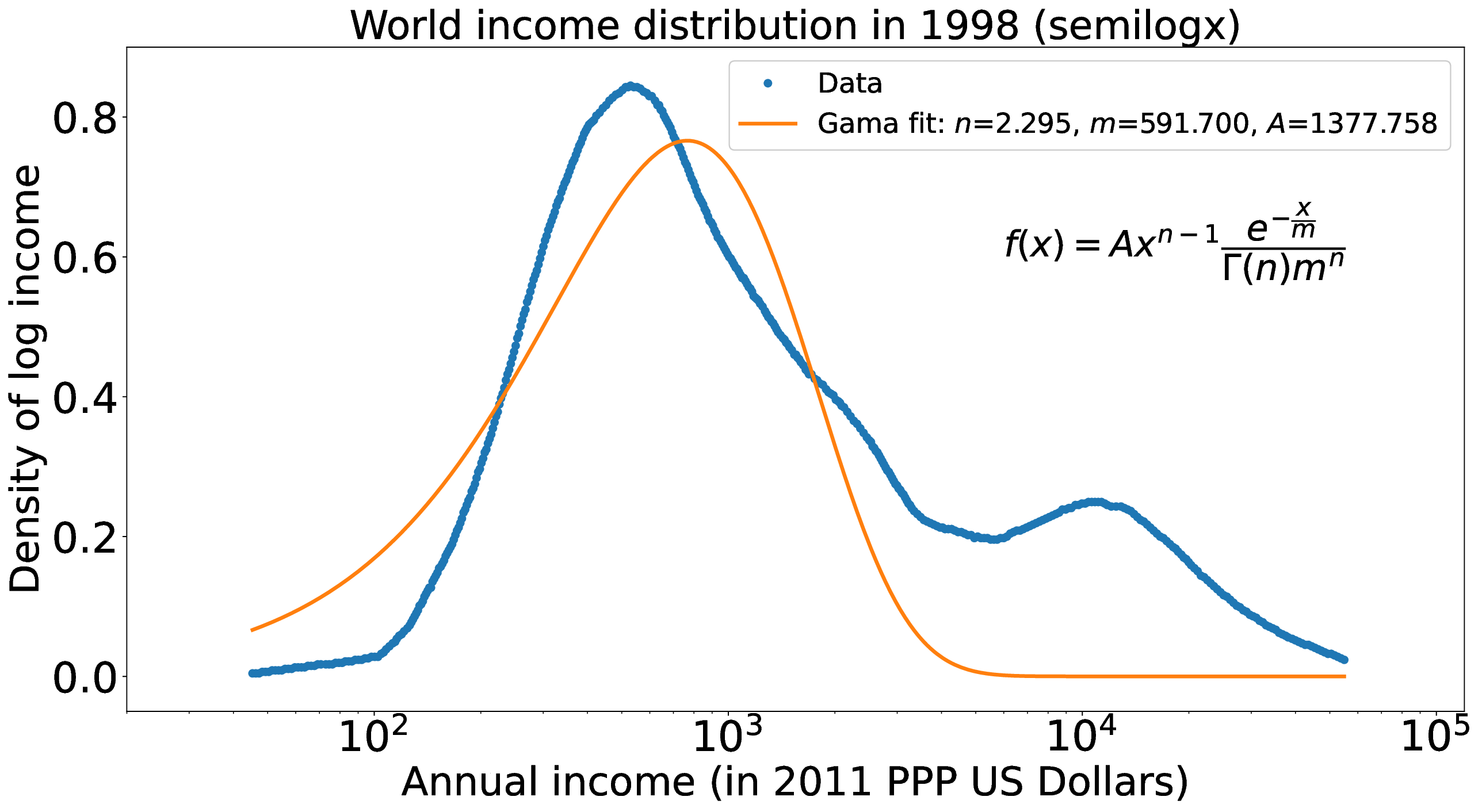}
    \caption{Gamma fit for year 1998, $R^2=0.72081$}\label{fig:Gama98}
\end{figure}

\begin{figure}[ht]
    \centering
    \includegraphics[width=\linewidth]{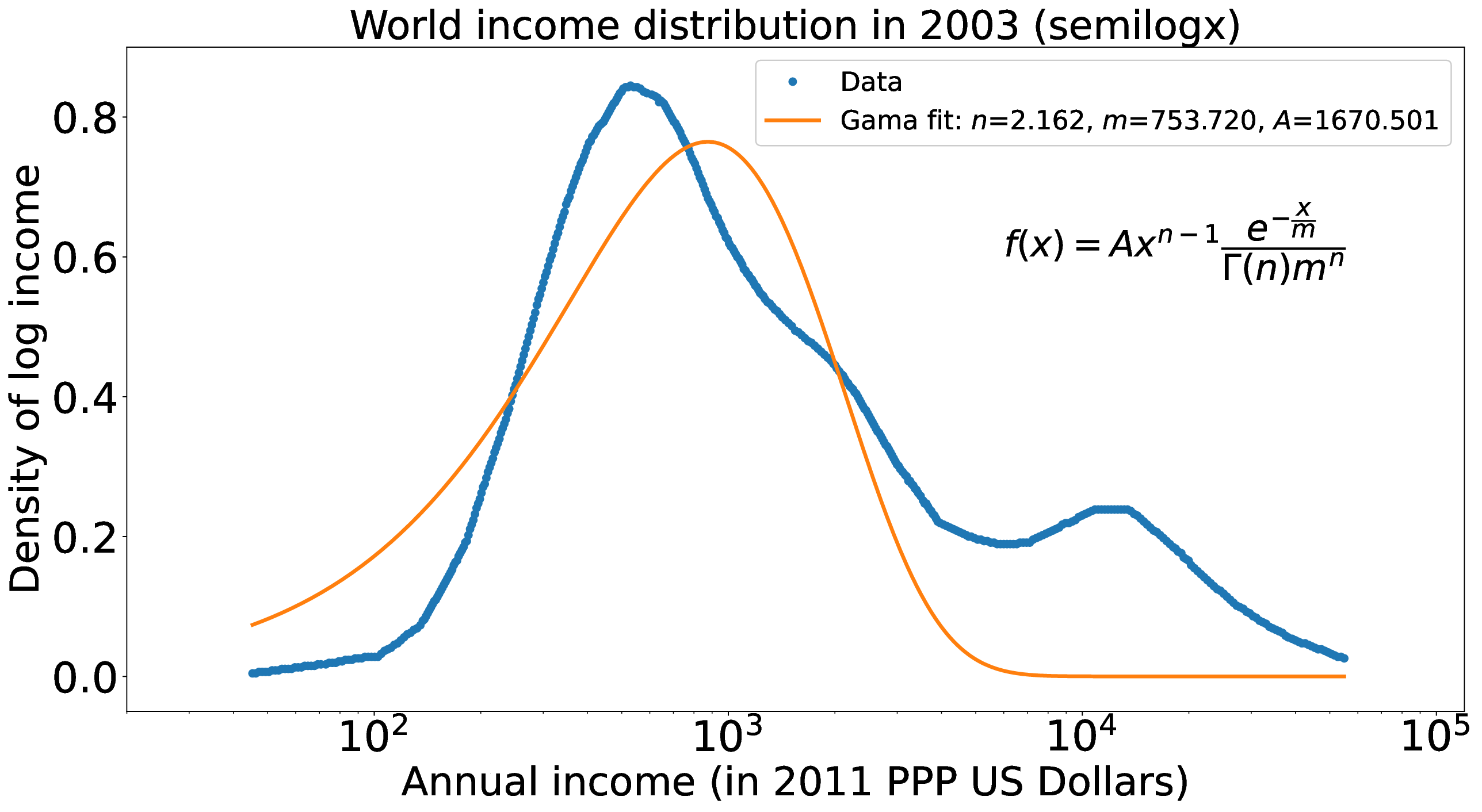}
    \caption{Gamma fit for year 2003, $R^2=0.73558$}\label{fig:Gama03}
\end{figure}

\begin{figure}[ht]
    \centering
    \includegraphics[width=\linewidth]{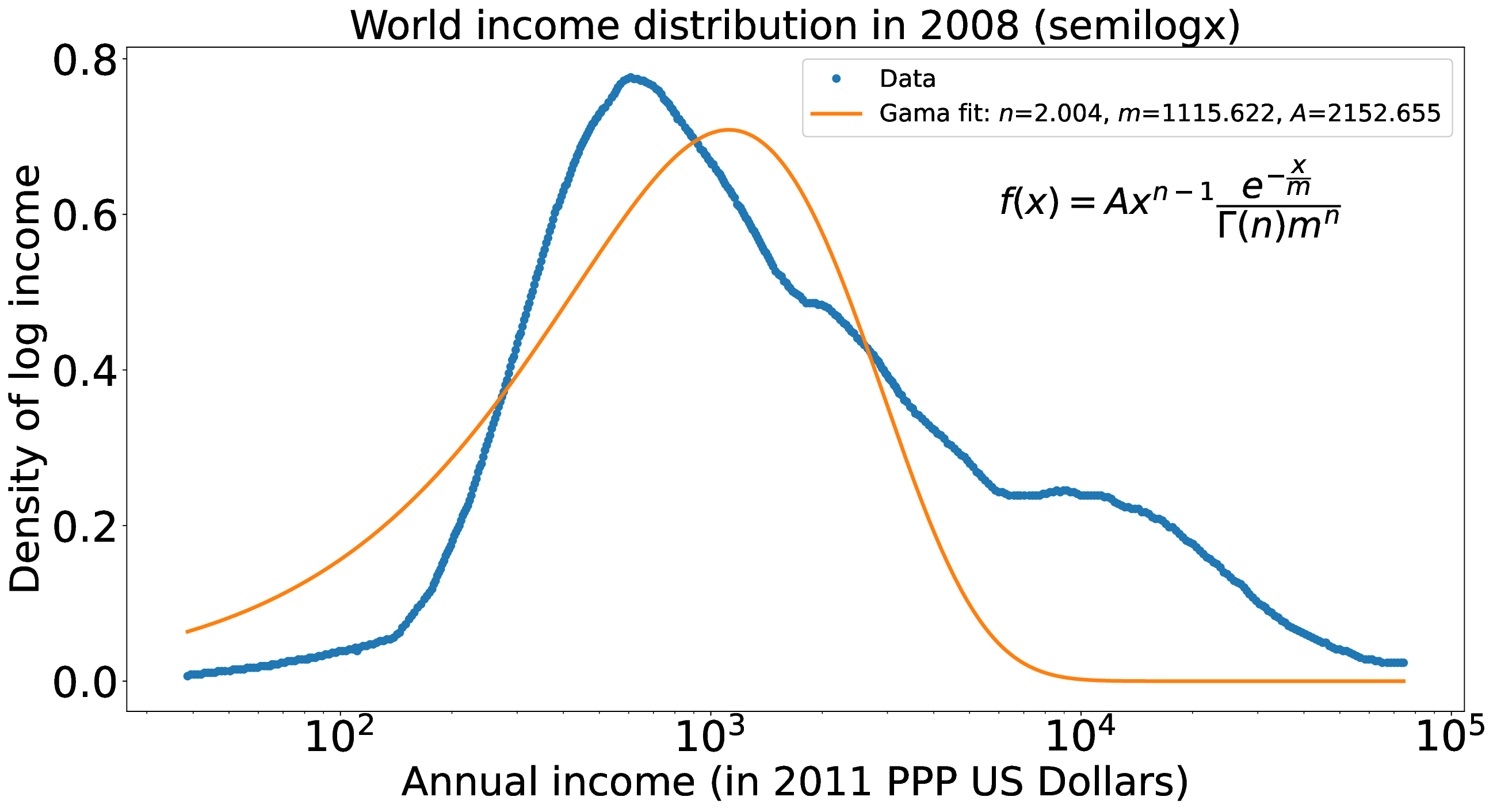}
    \caption{Gamma fit for year 2008, $R^2=0.72134$}\label{fig:Gama08}
\end{figure}

\begin{figure}[ht]
    \centering
    \includegraphics[width=\linewidth]{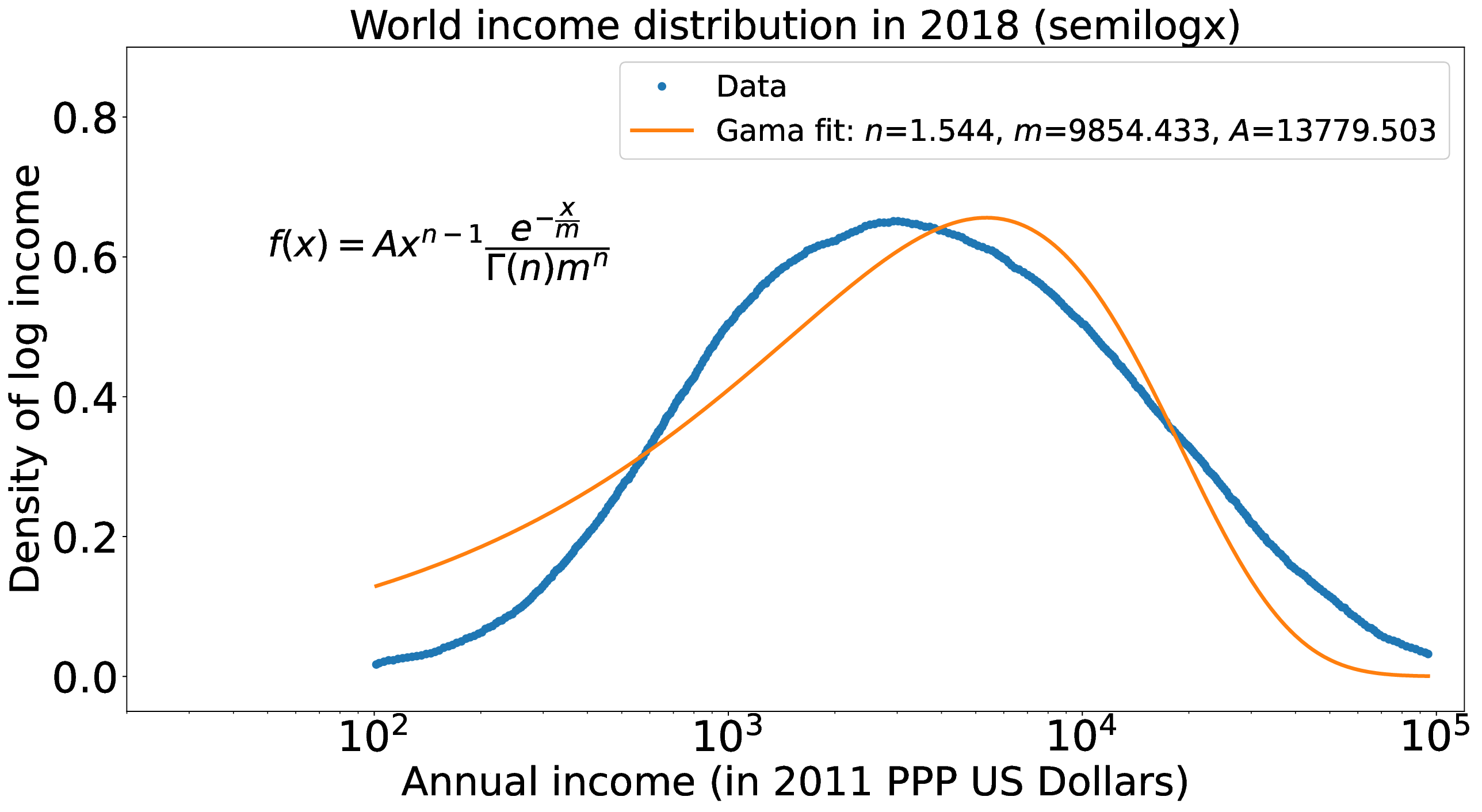}
    \caption{Gamma fit for year 2018, $R^2=0.85627$}\label{fig:Gama18}
\end{figure}

Figs.\ \ref{fig:Gama88} to \ref{fig:Gama18} show that a single gamma
distribution only matches the first peak. The adjustment parameter $R^2$
is around 0.72 except for year 2018 when it is 0.85. These plots also
show that a single gamma distribution only matches the first peak. The
adjustment parameter $R^2$ is around 0.72 except for year 2018 when it
is equal to 0.85. Table 1 shows the values of $R^2$ of all the fittings
we present here.

\begin{figure}[ht]
    \centering
    \includegraphics[width=\linewidth]{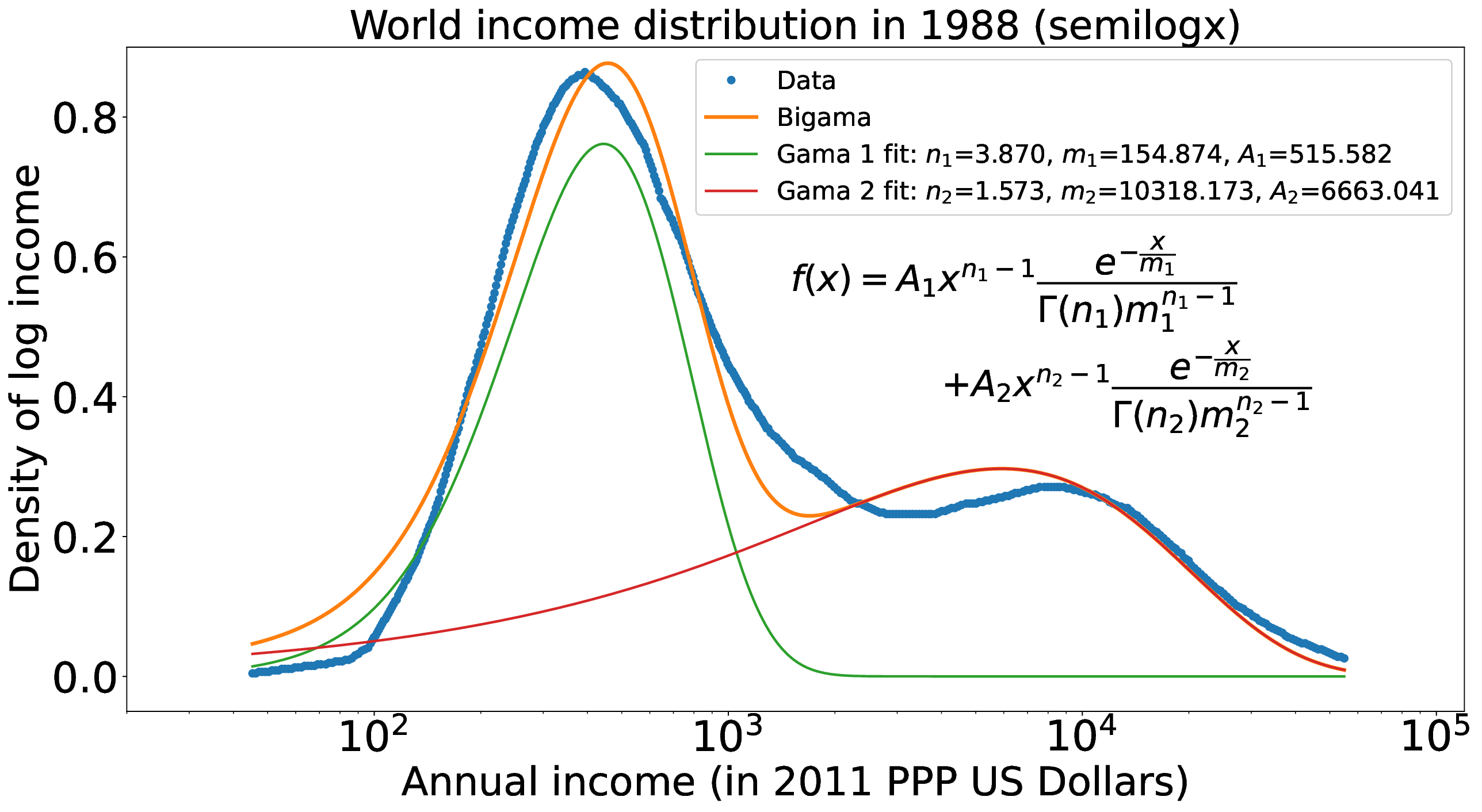}
    \caption{Bi-Gamma fit for Milanovic income distribution in 1988, $R^2=0.96115$}
    \label{fig:B88BG}
\end{figure}

\begin{figure}[ht]
    \centering
    \includegraphics[width=\linewidth]{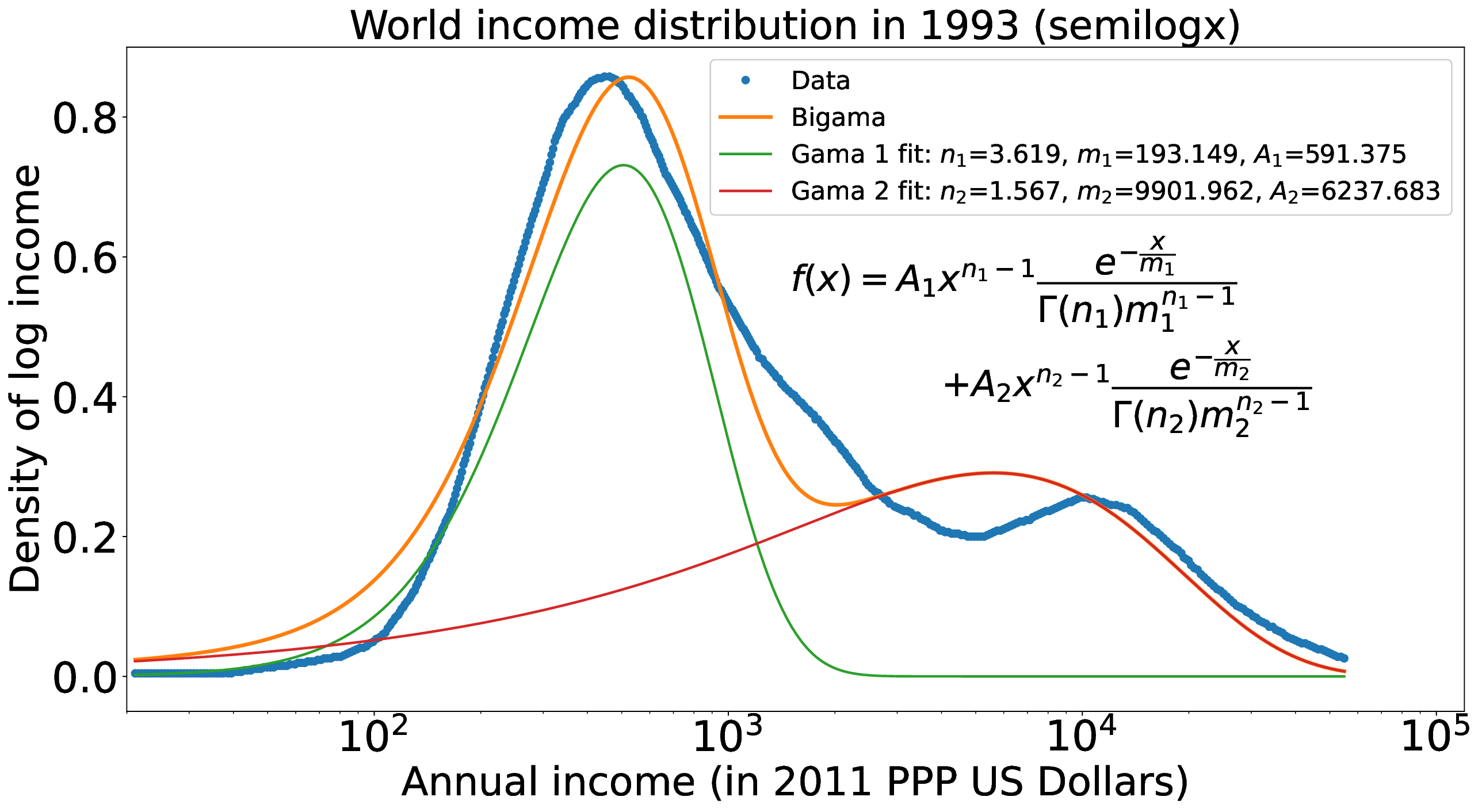}
    \caption{Same as Fig. \ref{fig:B88BG} but for 1993, $R^2=0.9499$}
    \label{fig:B93BG}
\end{figure}

\begin{figure}[ht]
    \centering
    \includegraphics[width=\linewidth]{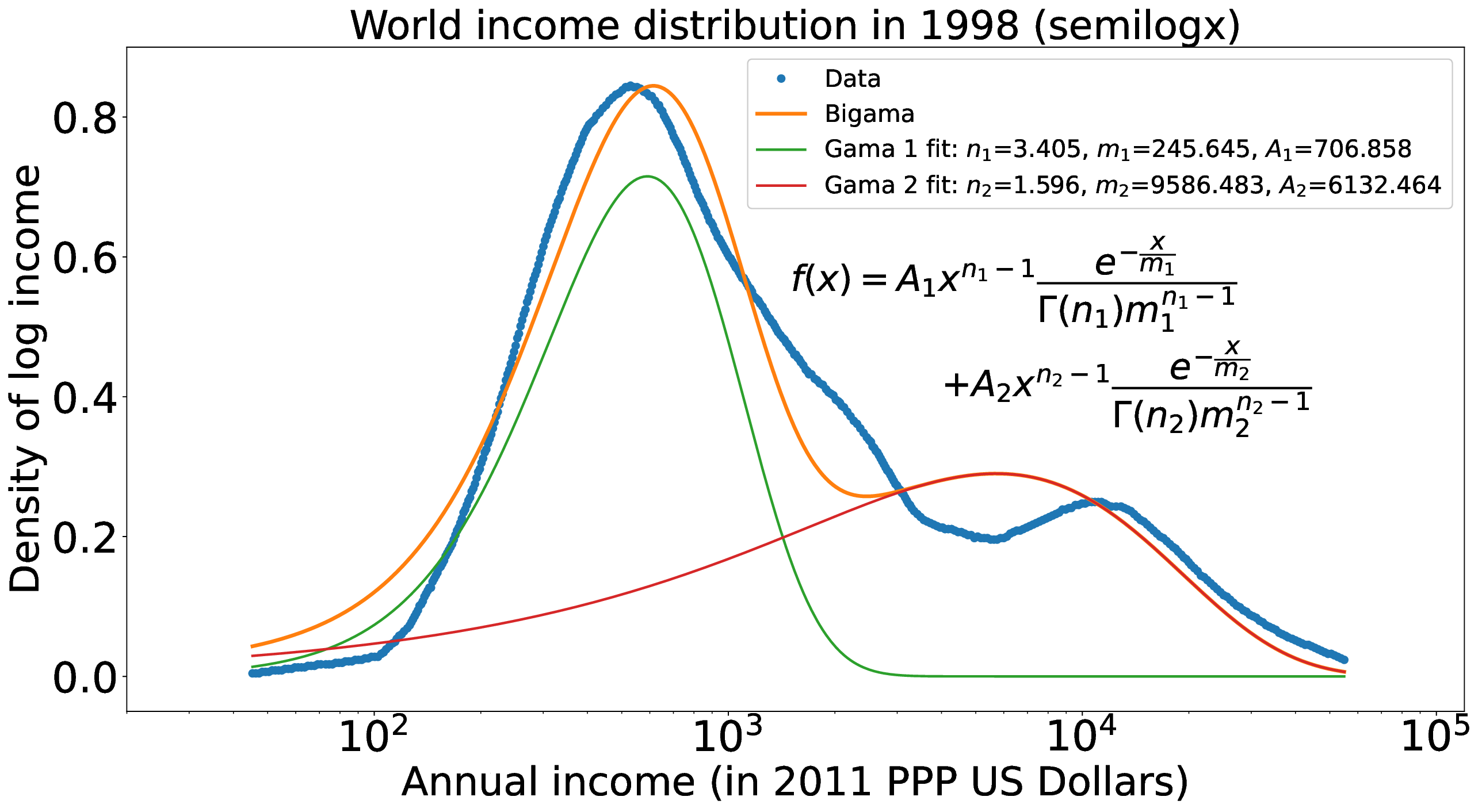}
    \caption{Same as Fig. \ref{fig:B88BG} but for 1998, $R^2=0.94064$}
    \label{fig:B98BG}
\end{figure}

\begin{figure}[ht]
    \centering
    \includegraphics[width=\linewidth]{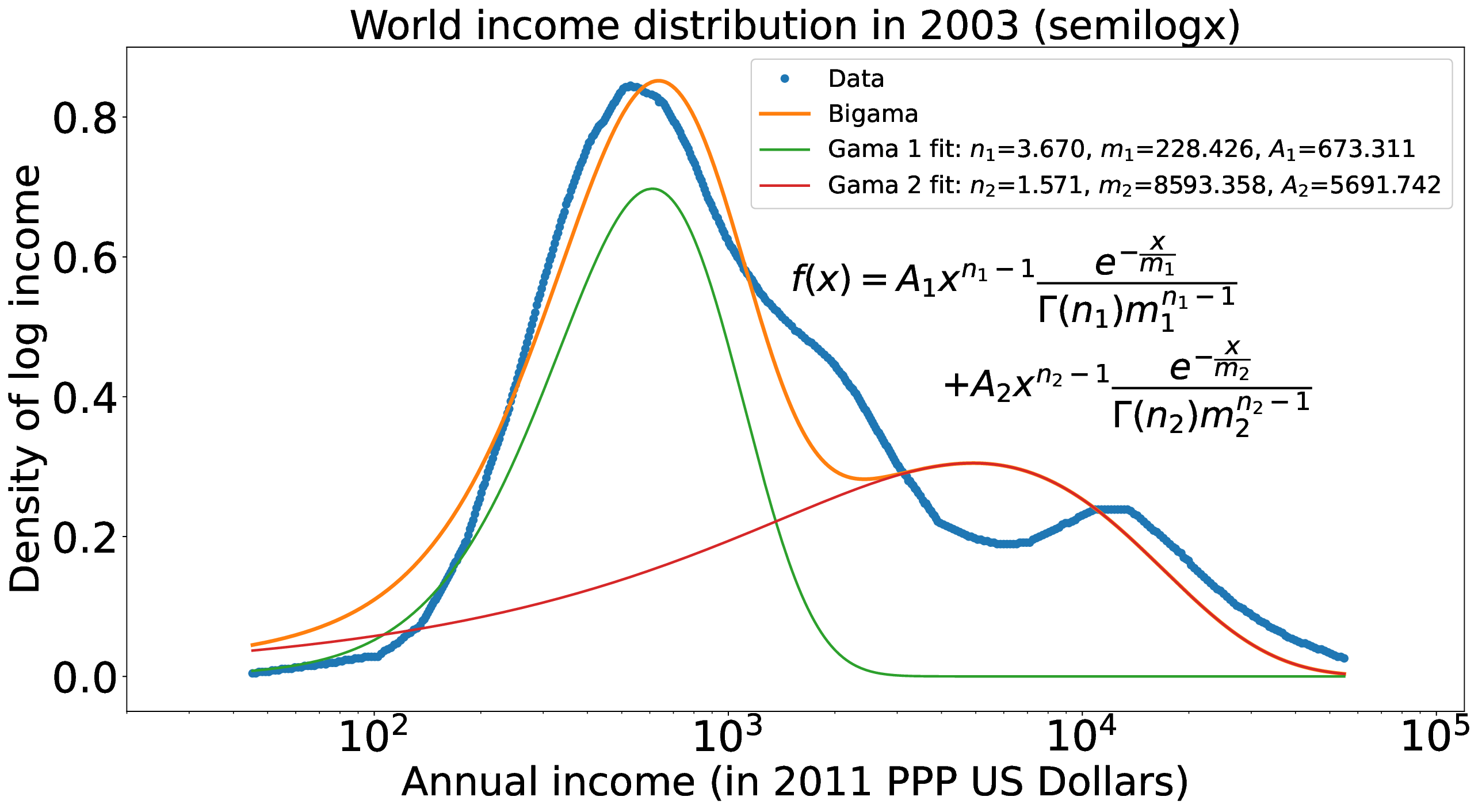}
    \caption{Same as Fig. \ref{fig:B88BG} but for 2003, $R^2=0.93372$}
    \label{fig:B03BG}
\end{figure}

\begin{figure}[ht]
    \centering
    \includegraphics[width=\linewidth]{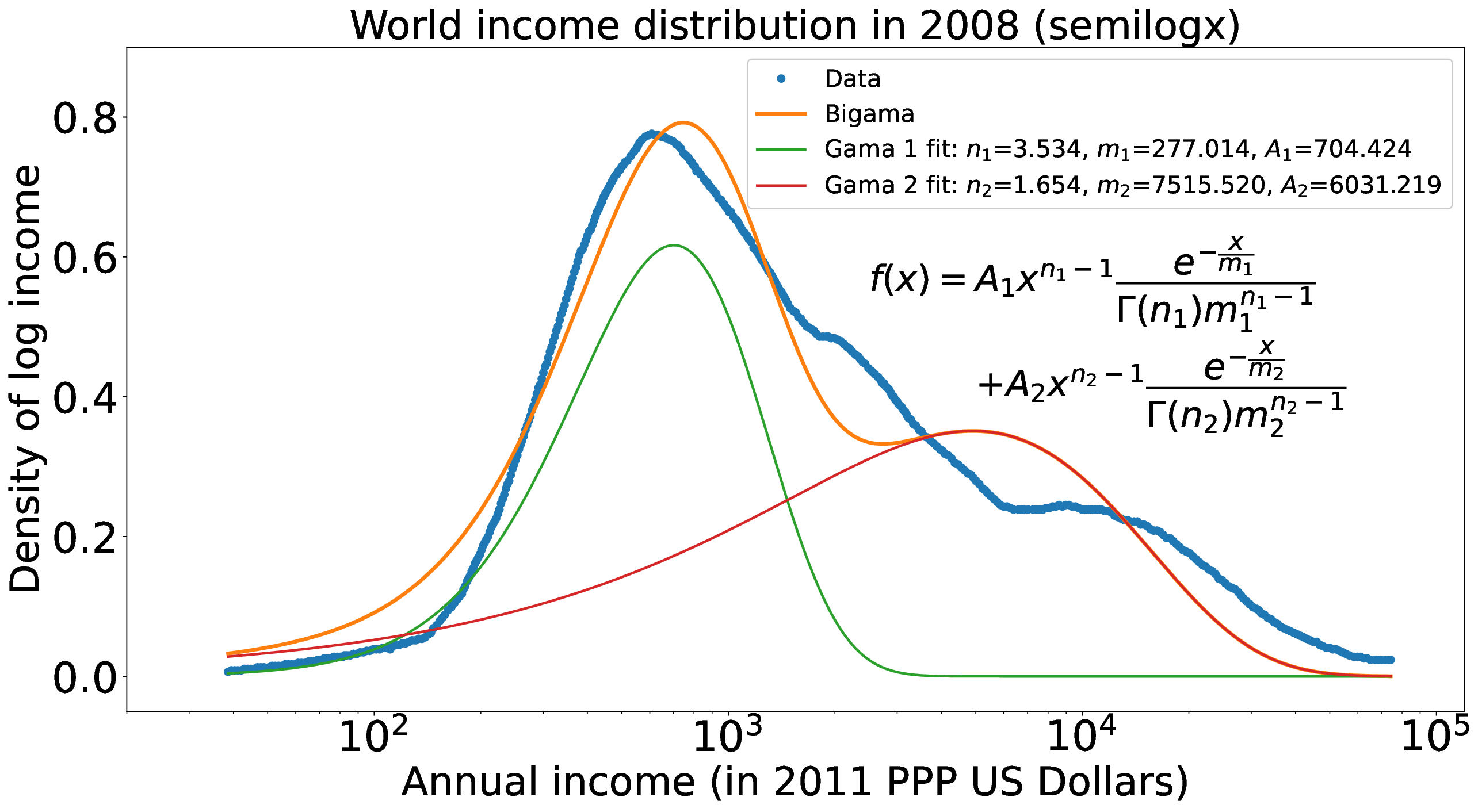}
    \caption{Same as Fig. \ref{fig:B88BG} but for 2008, $R^2=0.94681$}
    \label{fig:B08BG}
\end{figure}

\begin{figure}[ht]
    \centering
    \includegraphics[width=\linewidth]{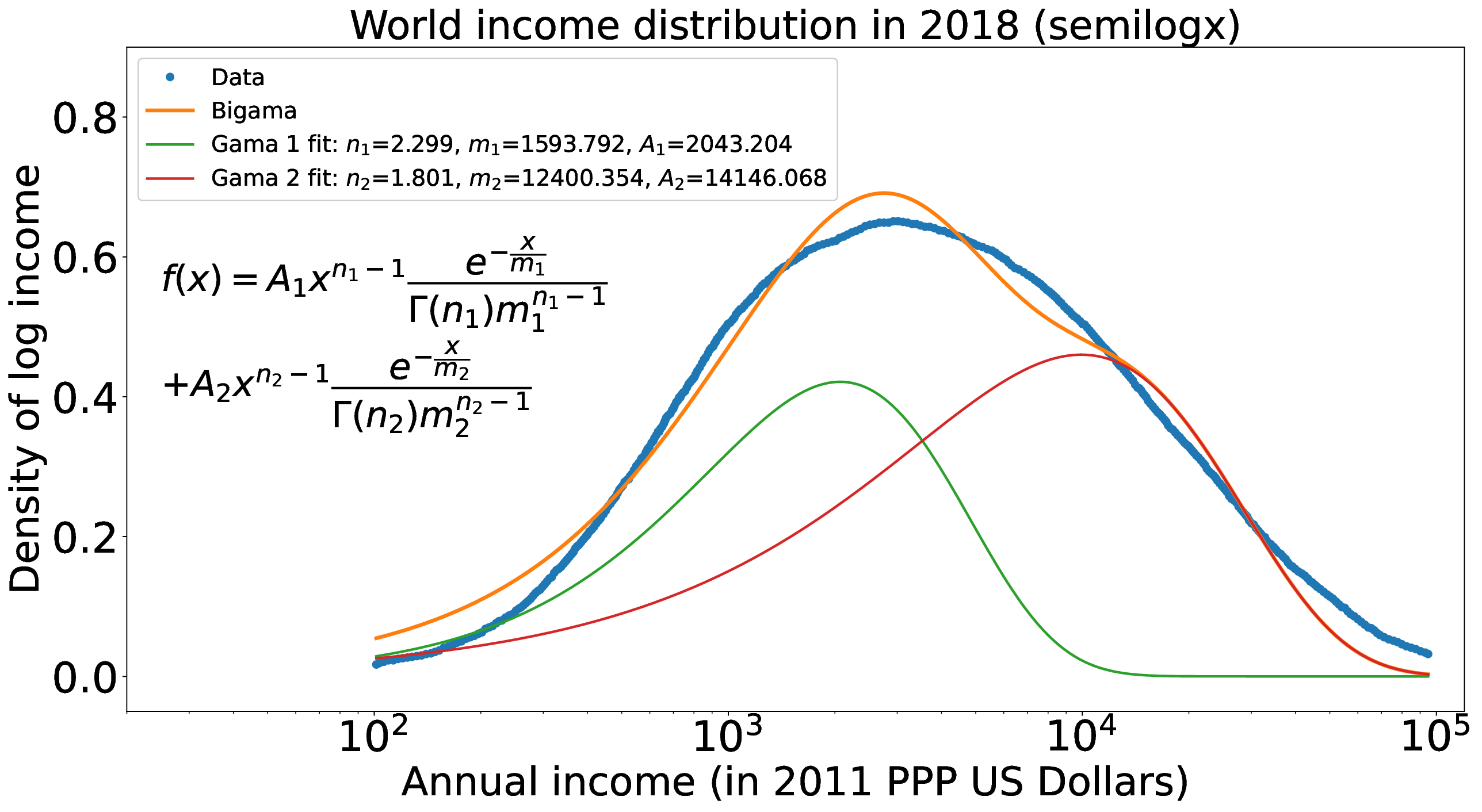}
    \caption{Same as Fig. \ref{fig:B88BG} but for 2018, $R^2=0.85627$}
    \label{fig:B18BG}
\end{figure}

Figs.\ \ref{fig:B88BG} to \ref{fig:B18BG} show basically the same data as
in previous figures now fitted with bi-gamma functions, either separately
or together. The plots show that each one fits well some portion of the
data and that taking them together provides a better fit to the whole
distribution. In general the fitting is better in the low ``poor'' region
than in the ``rich'' one.

\begin{figure}[ht]
    \centering
    \includegraphics[width=\linewidth]{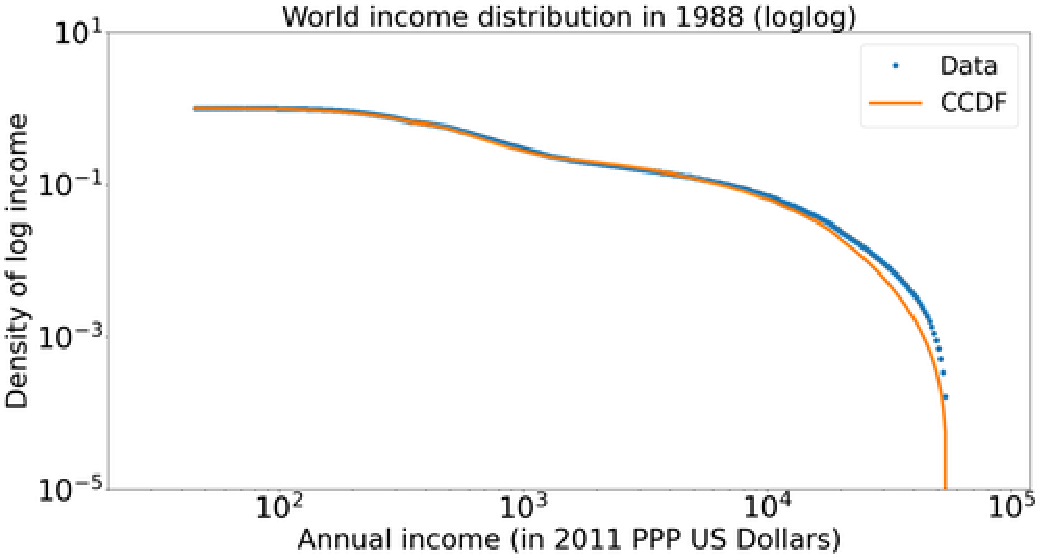}
    \caption{CCDF bi-gamma for 1988, $R^2=0.99782$}
    \label{fig:BC88}
\end{figure}

\begin{figure}[ht]
    \centering
    \includegraphics[width=\linewidth]{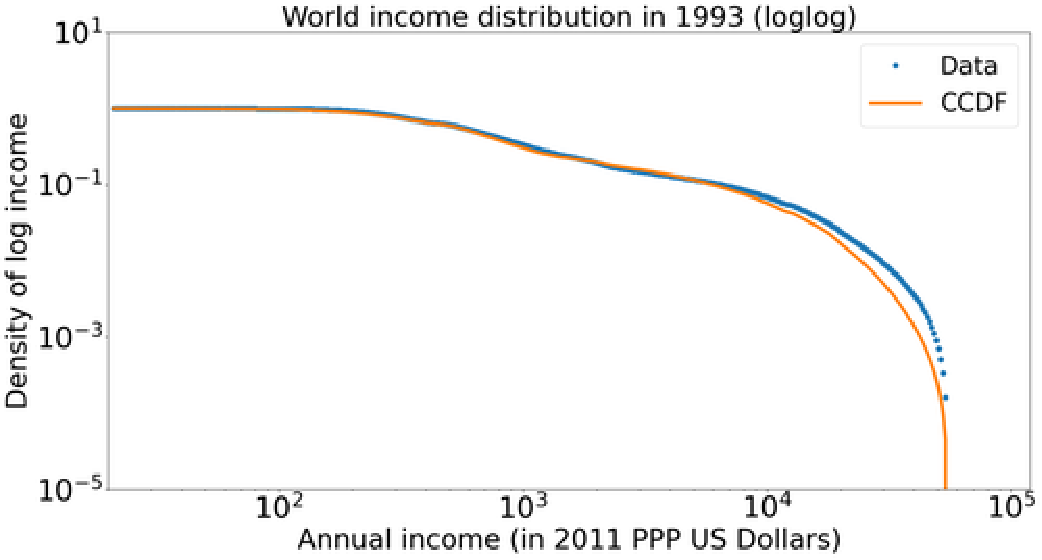}
    \caption{CCDF bi-gamma for 1993, $R^2=0.99712$}
    \label{fig:BC93}
\end{figure}

\begin{figure}[ht]
    \centering
    \includegraphics[width=\linewidth]{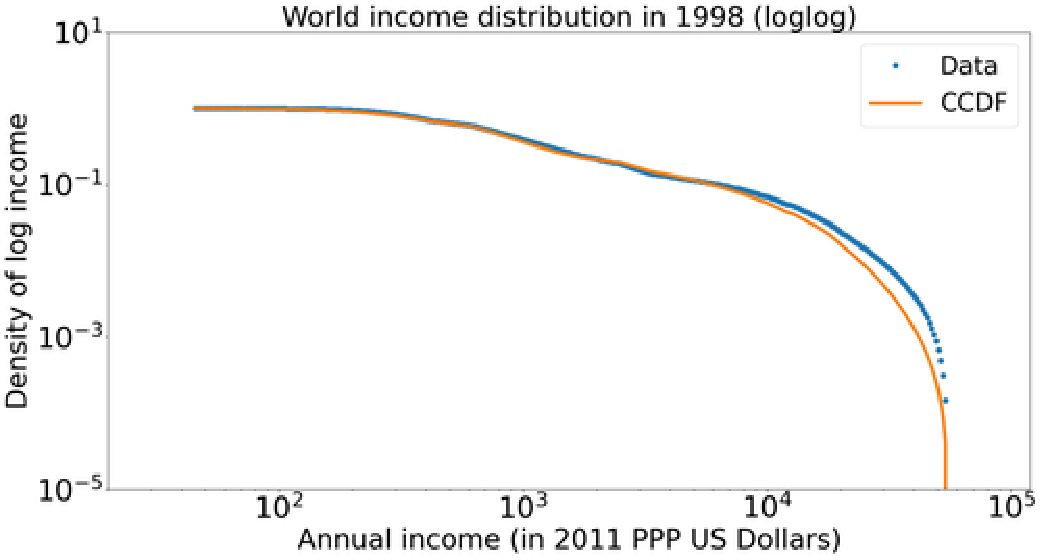}
    \caption{CCDF bi-gamma for 1998, $R^2=0.99721$}
    \label{fig:BC98}
\end{figure}

\begin{figure}[ht]
    \centering
    \includegraphics[width=\linewidth]{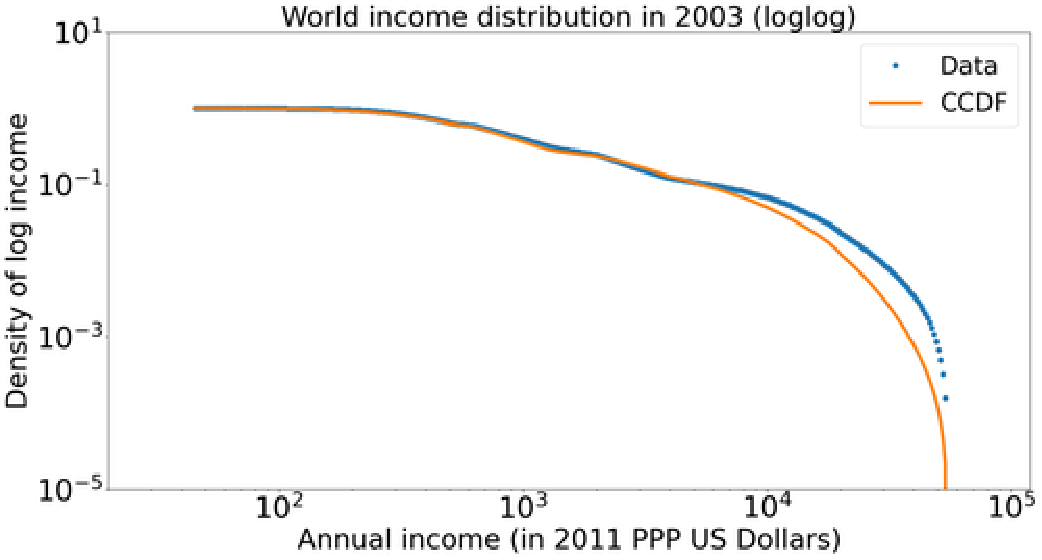}
    \caption{CCDF bi-gamma for 2003, $R^2=0.99711$}
    \label{fig:BC03}
\end{figure}

\begin{figure}[ht]
    \centering
    \includegraphics[width=\linewidth]{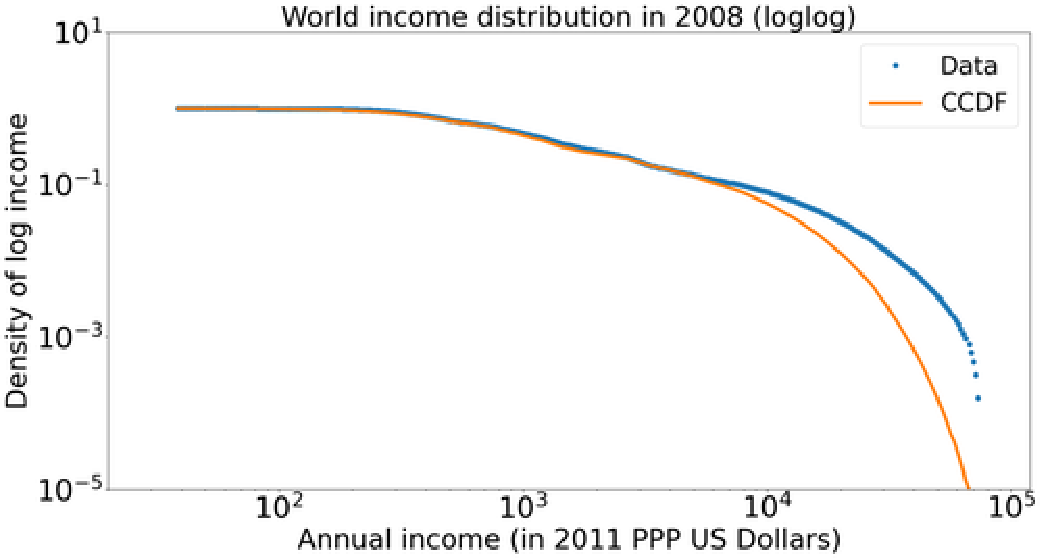}
    \caption{CCDF bi-gamma for 2008, $R^2=0.99685$}
    \label{fig:BC08}
\end{figure}

\begin{figure}[ht]
    \centering
    \includegraphics[width=\linewidth]{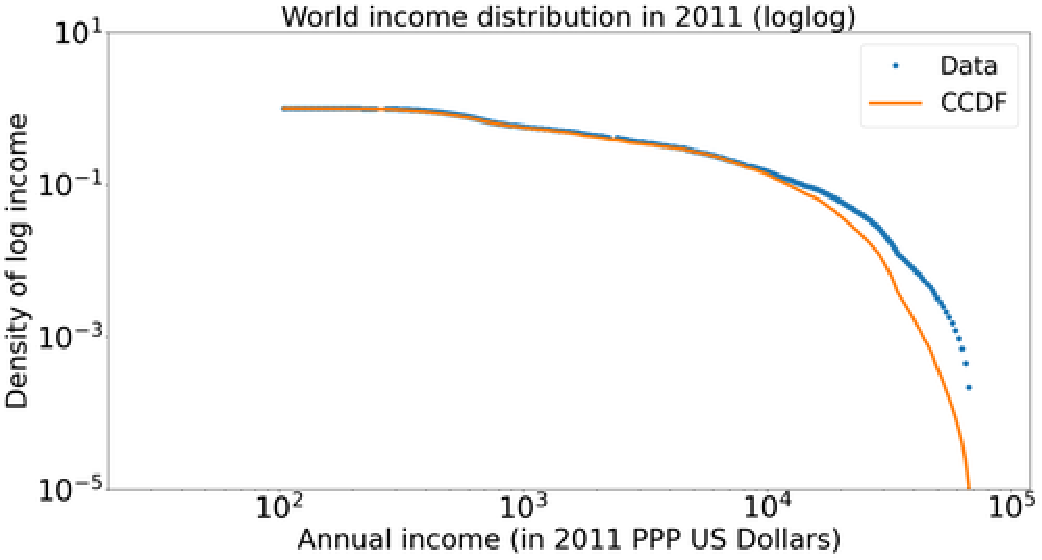}
    \caption{CCDF bi-gamma for 2011, $R^2=0.99743$}
    \label{fig:BC11}
\end{figure}

\begin{figure}[ht]
    \centering
    \includegraphics[width=\linewidth]{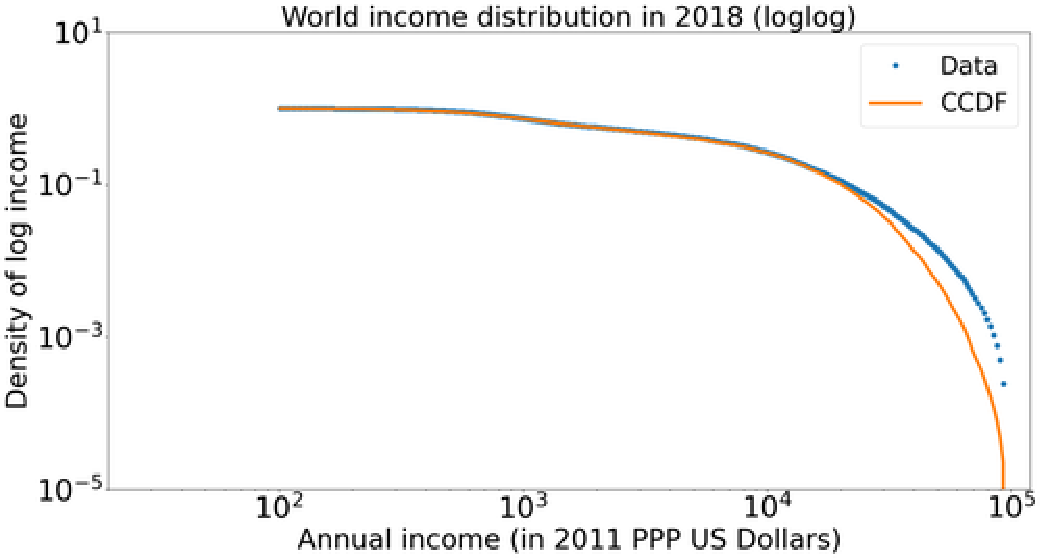}
    \caption{CCDF bi-gamma for 2018, $R^2=0.99896$}
    \label{fig:BC18}
\end{figure}

Figs.\ \ref{fig:BC88} to \ref{fig:BC18} present the same data as in
previous figures, but fitted to the CCDF bi-gamma to obtain better
values for $R^2$ than the PDF fittings. It is clear how in the bi-gamma
CCDF curves deviate slightly below the ones of the empirical values at
the tail of the distribution, that is, in the rich region.

\subsection{Log-normal and bi-log-normal fits}

\begin{figure}[ht]
    \centering
    \includegraphics[width=\linewidth]{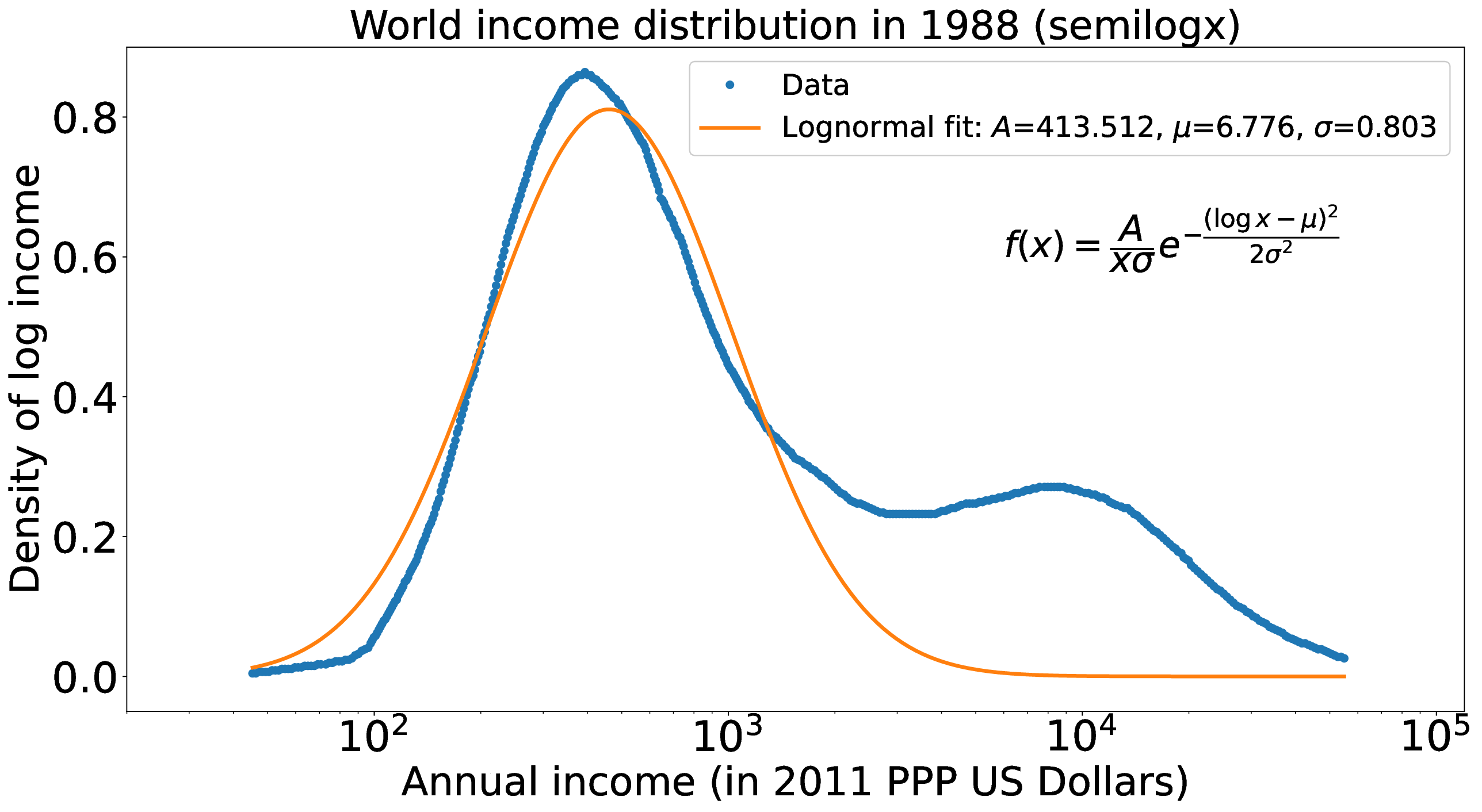}
    \caption{Log-normal fit for Milanovic income distribution
in 1988, $R^2=0.77358$}
    \label{fig:BLN88}
\end{figure}

\begin{figure}[ht]
    \centering
    \includegraphics[width=\linewidth]{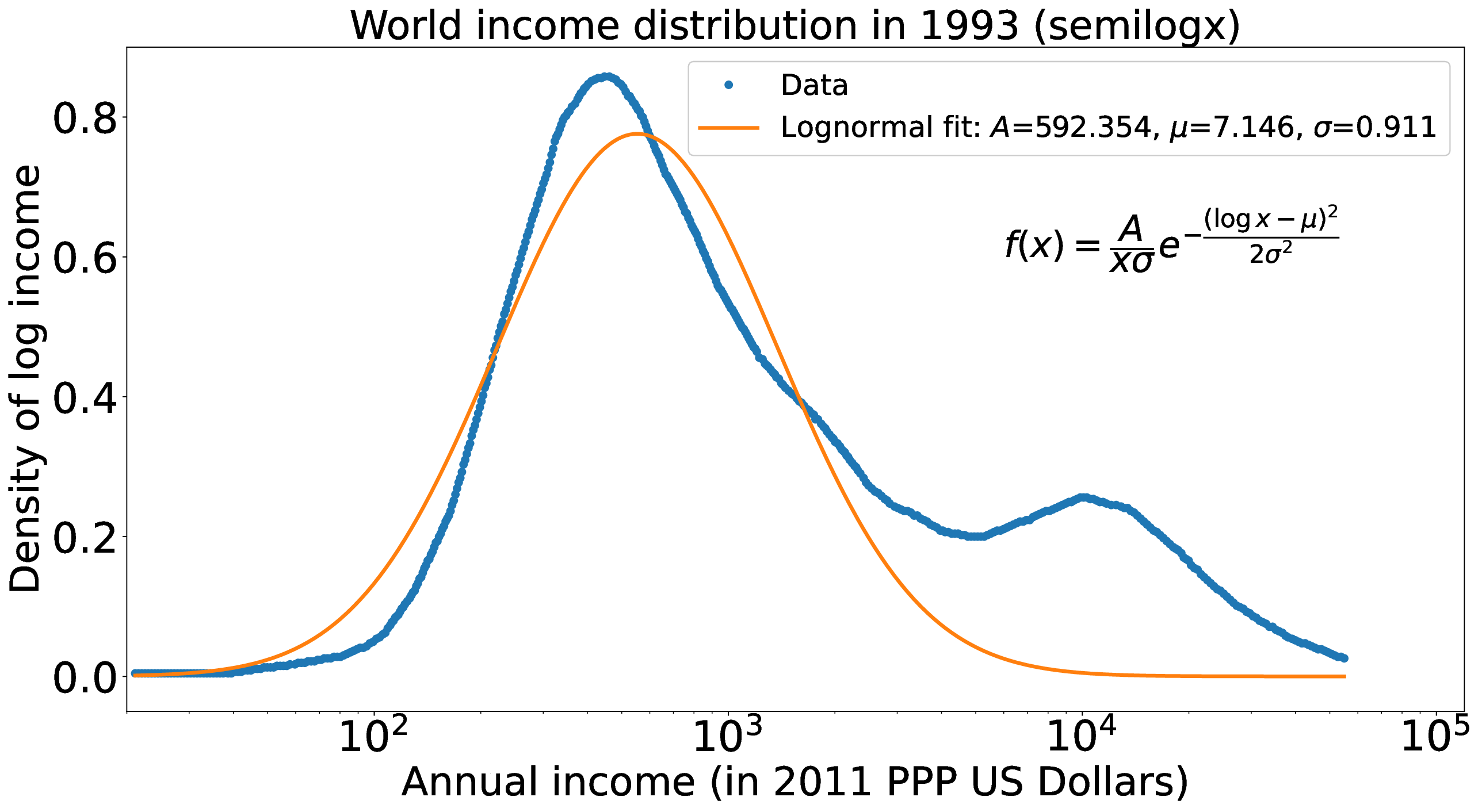}
    \caption{Same as Fig. \ref{fig:BLN88} but for 1993, $R^2=0.8366$}
    \label{fig:BLN93}
\end{figure}

\begin{figure}[ht]
    \centering
    \includegraphics[width=\linewidth]{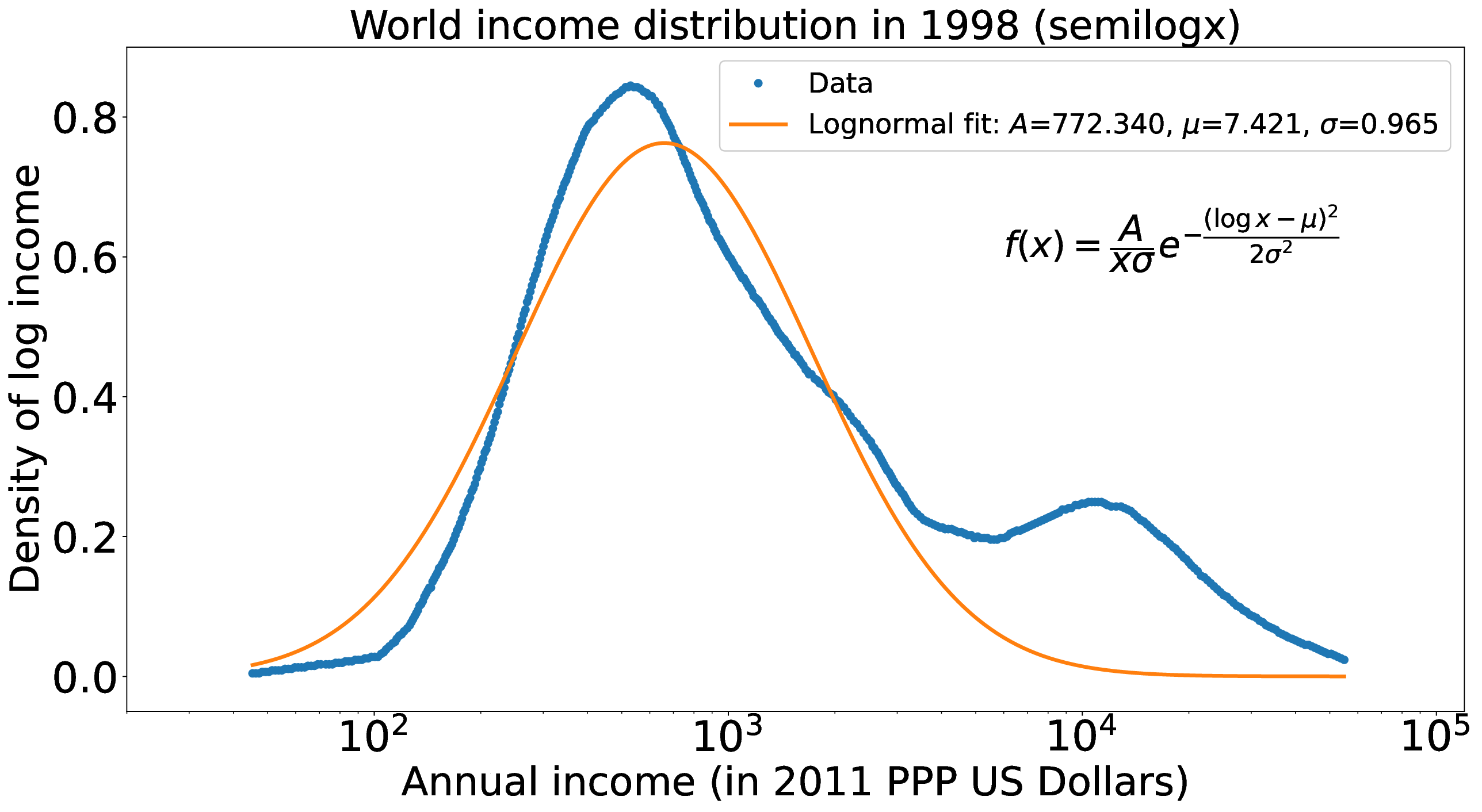}
    \caption{Same as Fig. \ref{fig:BLN88} but for 1998, $R^2=0.84092$}
    \label{fig:BLN98}
\end{figure}

\begin{figure}[ht]
    \centering
    \includegraphics[width=\linewidth]{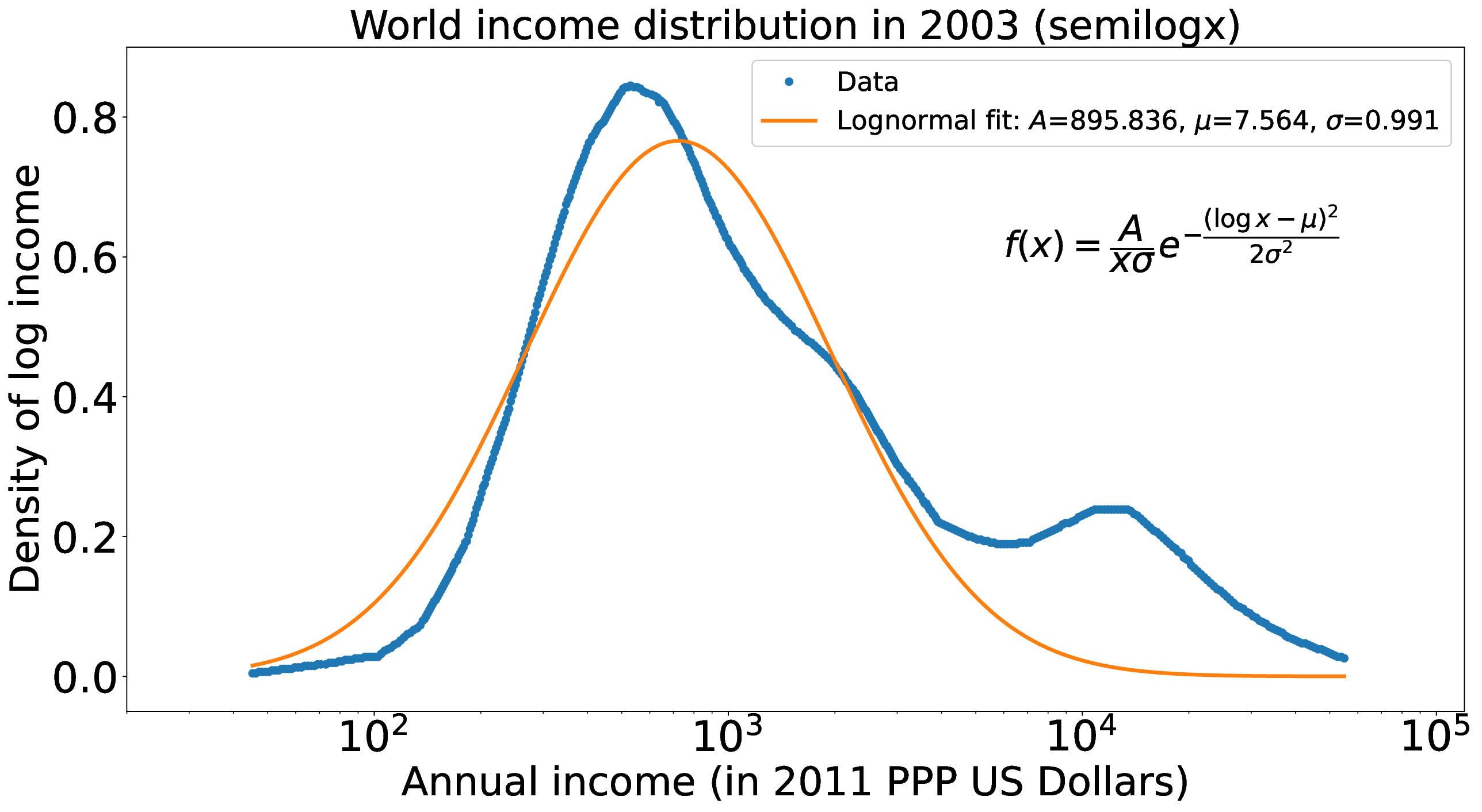}
    \caption{Same as Fig. \ref{fig:BLN88} but for 2003, $R^2=0.85684$}
    \label{fig:BLN03}
\end{figure}

\begin{figure}[ht]
    \centering
    \includegraphics[width=\linewidth]{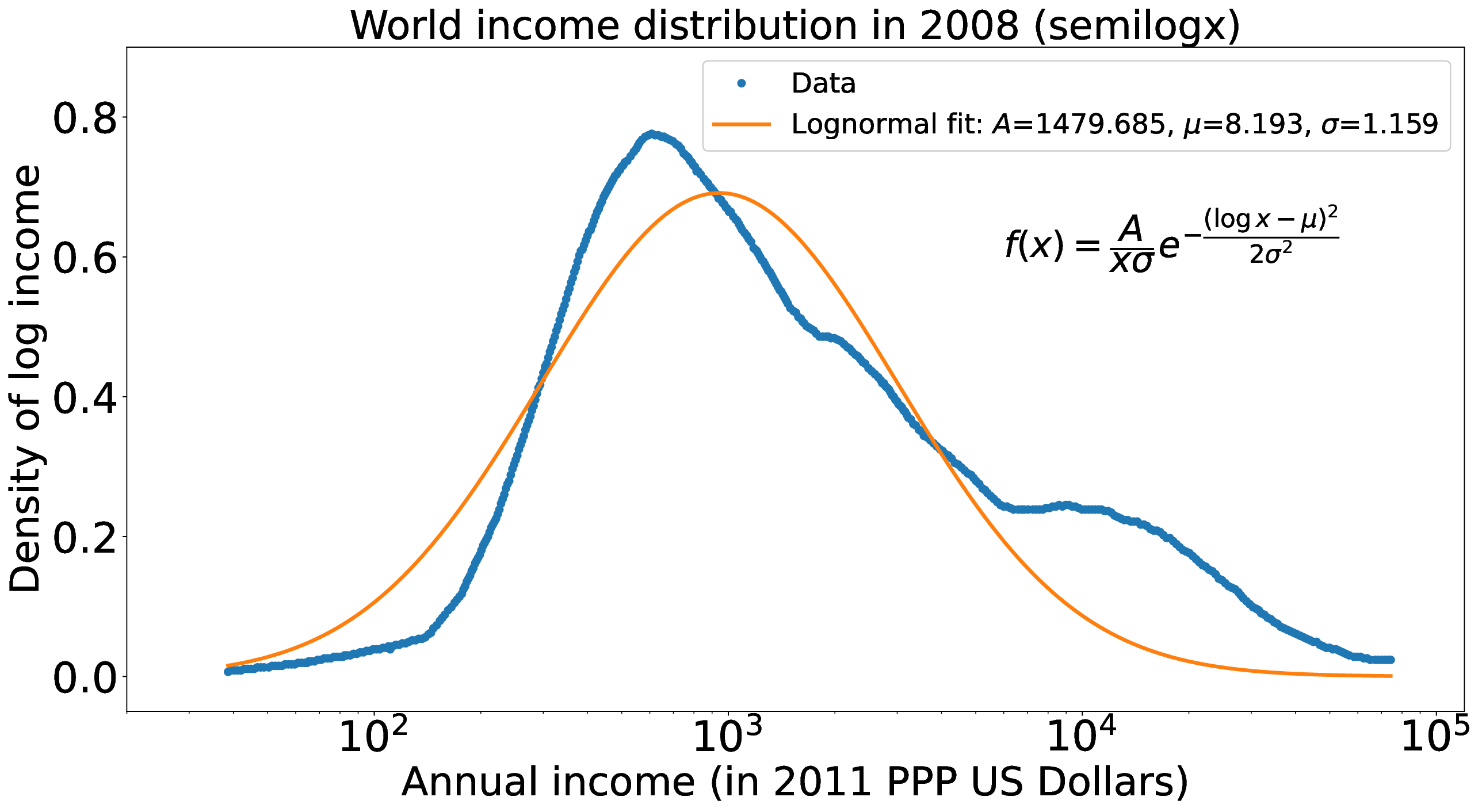}
    \caption{Same as Fig. \ref{fig:BLN88} but for 2008, $R^2=0.86455$}
    \label{fig:BLN033}
\end{figure}

\begin{figure}[ht]
    \centering
    \includegraphics[width=\linewidth]{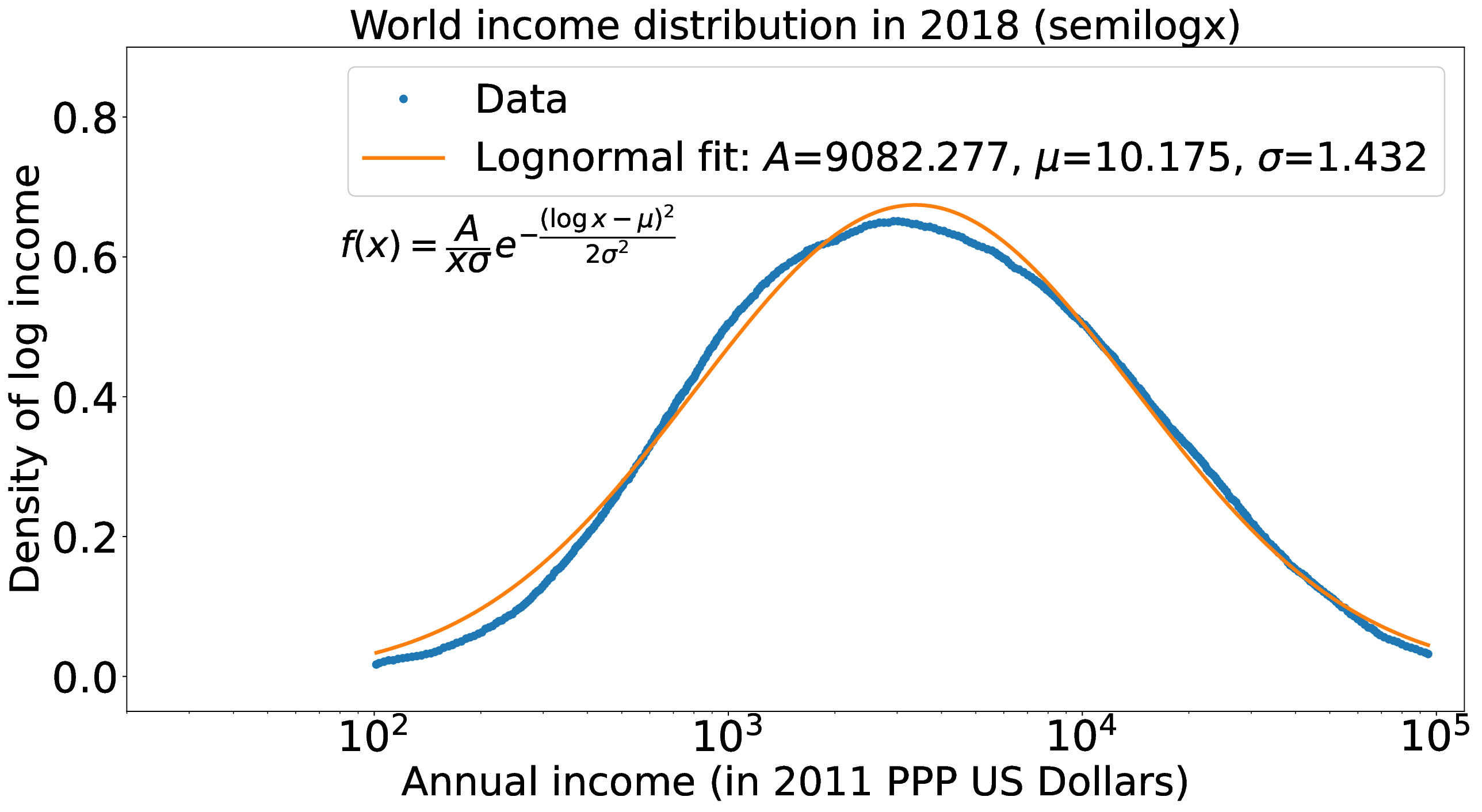}
    \caption{Same as Fig. \ref{fig:BLN88} but for 2018, $R^2=0.98991$}
    \label{fig:BLN18}
\end{figure}

\begin{figure}[ht]
    \centering
    \includegraphics[width=\linewidth]{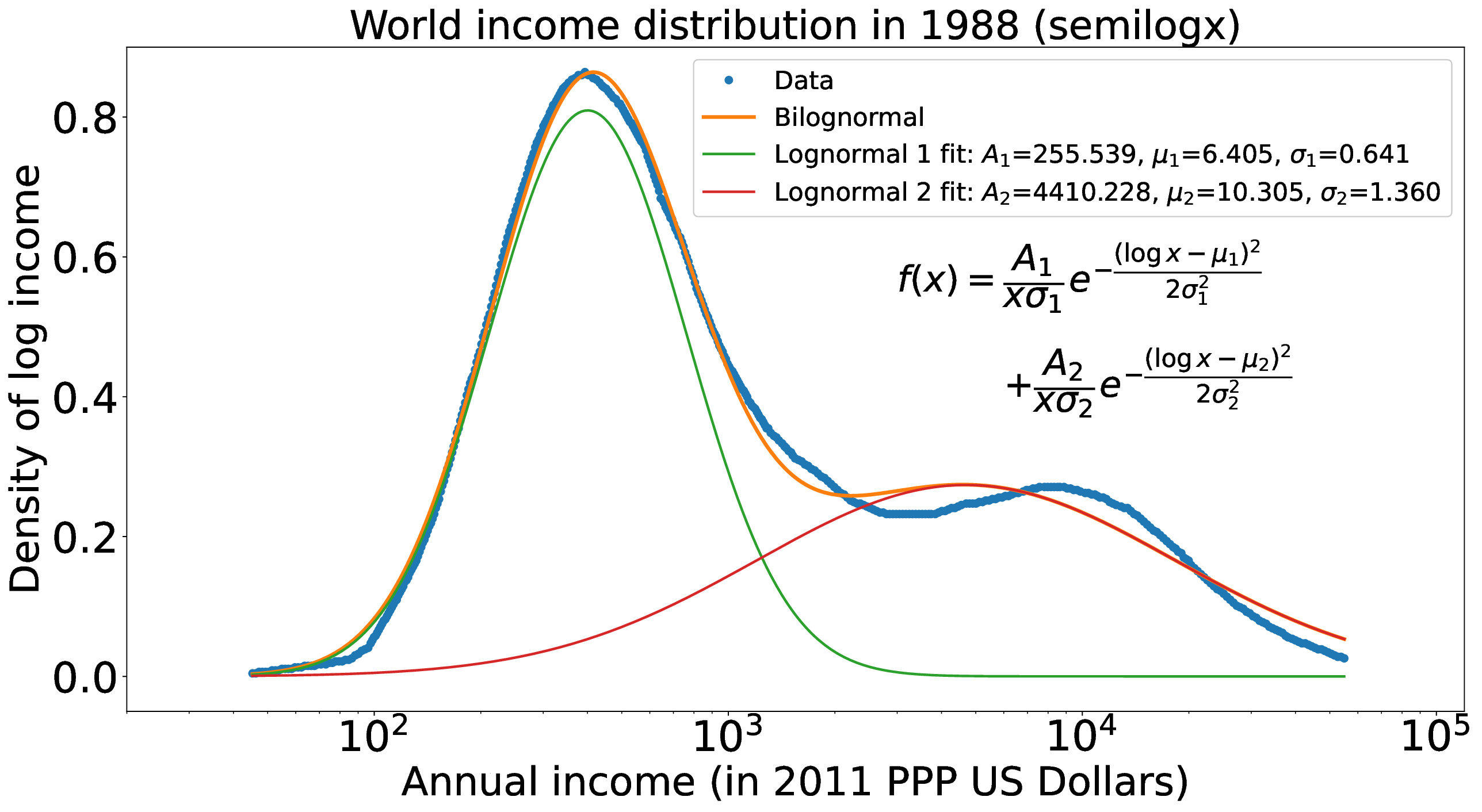}
    \caption{Bi-log-normal fit for Milanovic income distribution
in 1988, $R^2=0.99383$.}
    \label{fig:B88BL}
\end{figure}

\begin{figure}[ht]
    \centering
    \includegraphics[width=\linewidth]{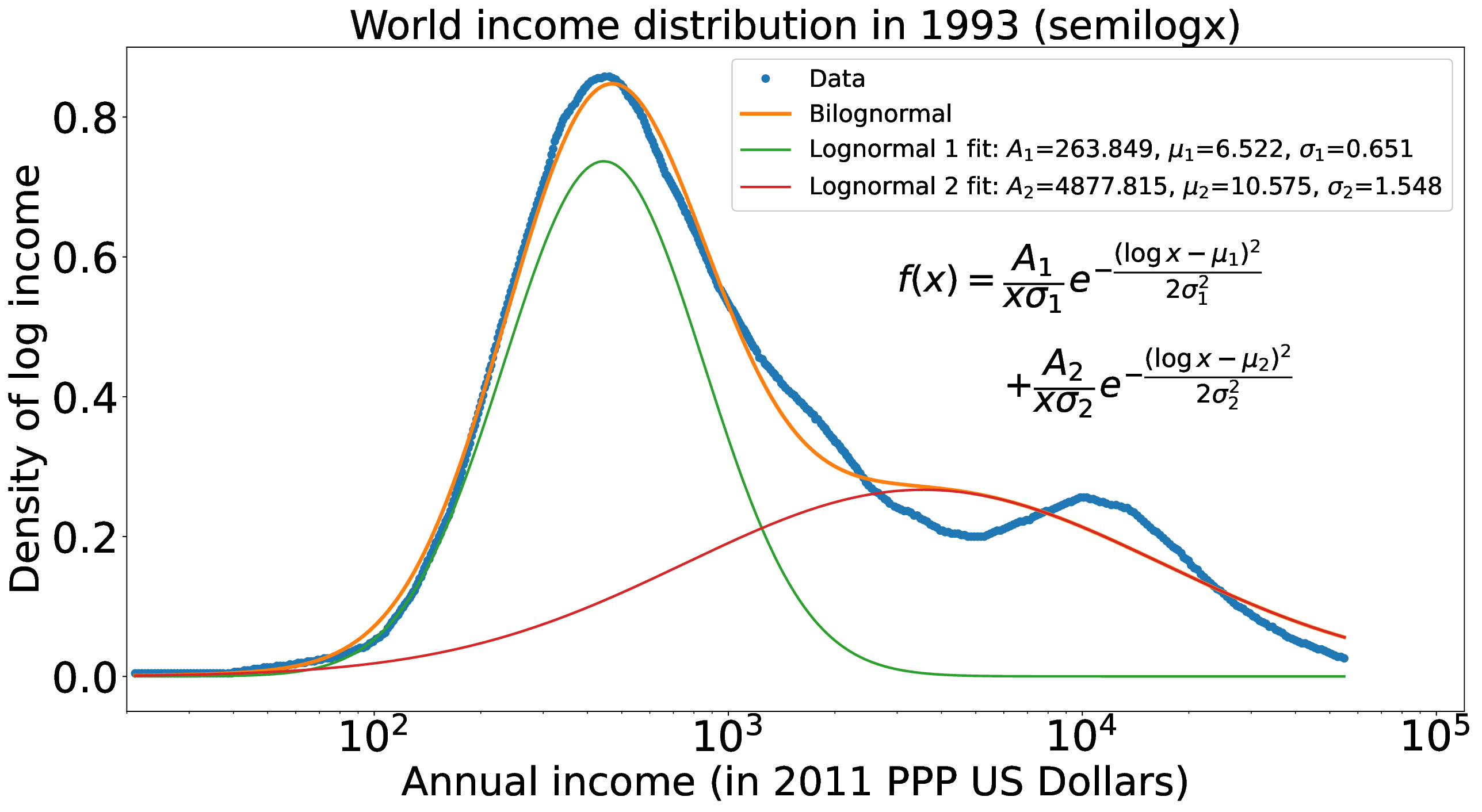}
    \caption{Same as Fig. \ref{fig:B88BL} but for 1993 $R^2=0.98898$.}
    \label{fig:B93BL}
\end{figure}

\begin{figure}[ht]
    \centering
    \includegraphics[width=\linewidth]{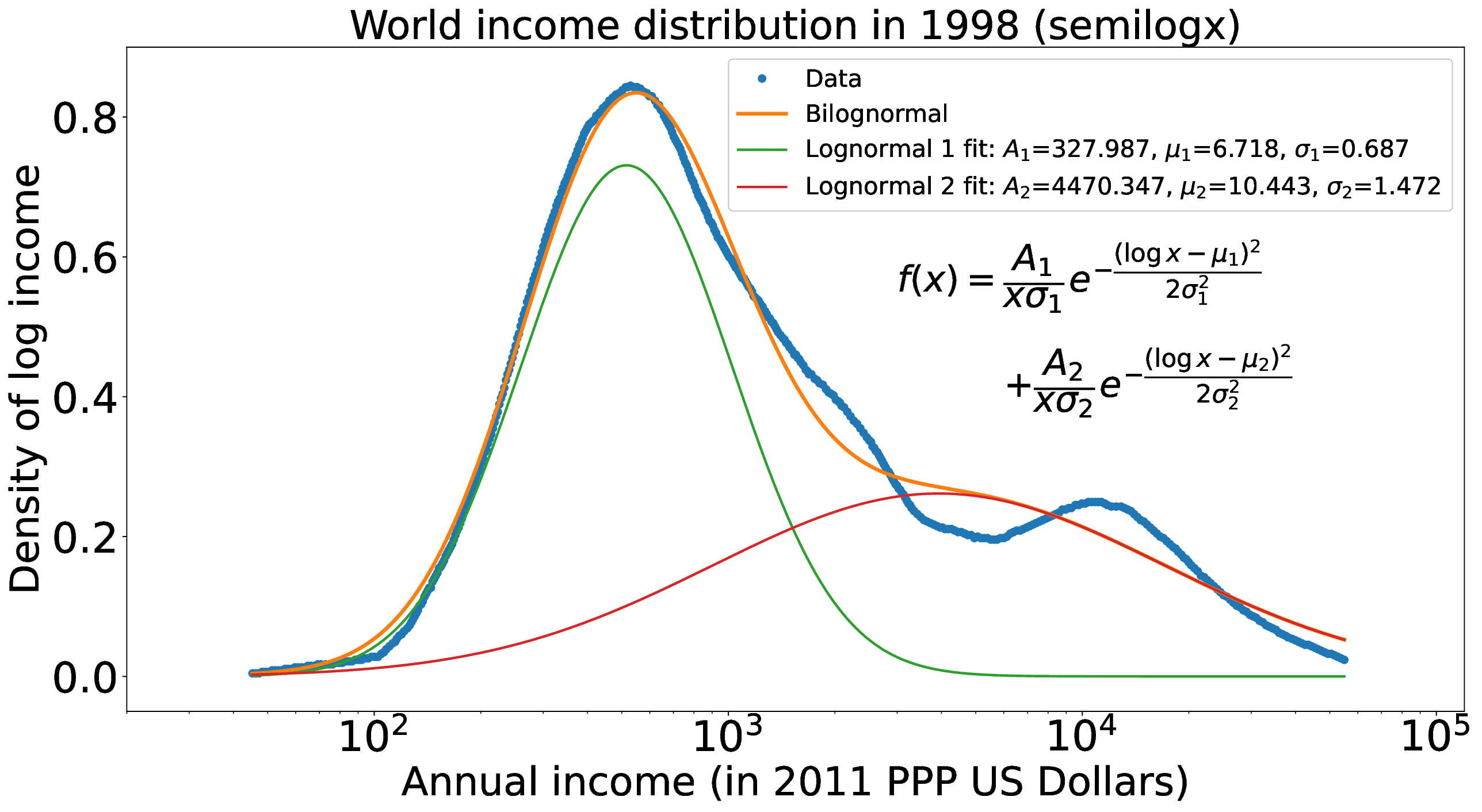}
    \caption{Same as Fig. \ref{fig:B88BL} but for 1998 $R^2=0.98634$}
    \label{fig:B98BL}
\end{figure}

\begin{figure}[ht]
    \centering
    \includegraphics[width=\linewidth]{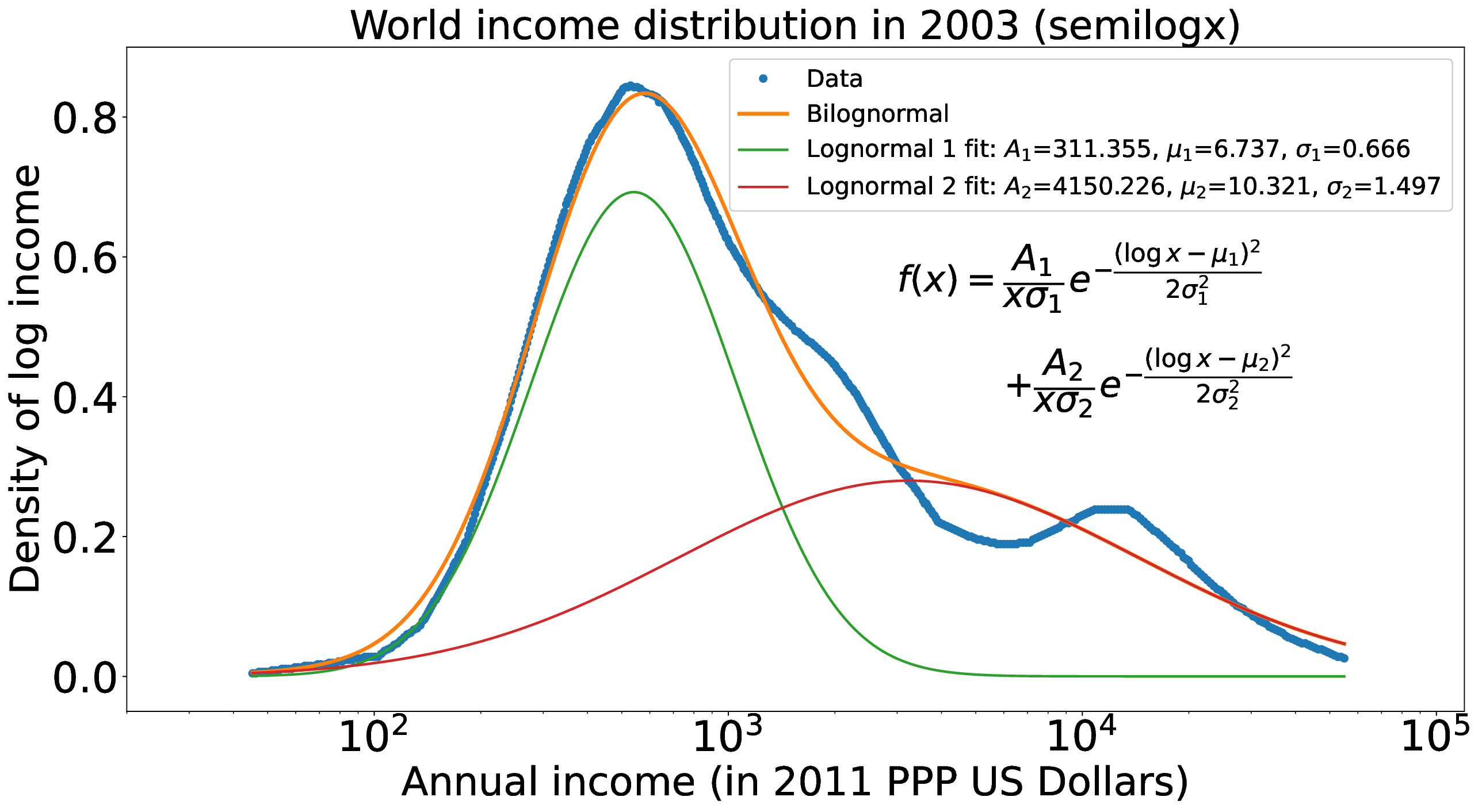}
    \caption{Same as Fig. \ref{fig:B88BL} but for 2003 $R^2=0.98224$}
    \label{fig:B03BL}
\end{figure}

\begin{figure}[ht]
    \centering
    \includegraphics[width=\linewidth]{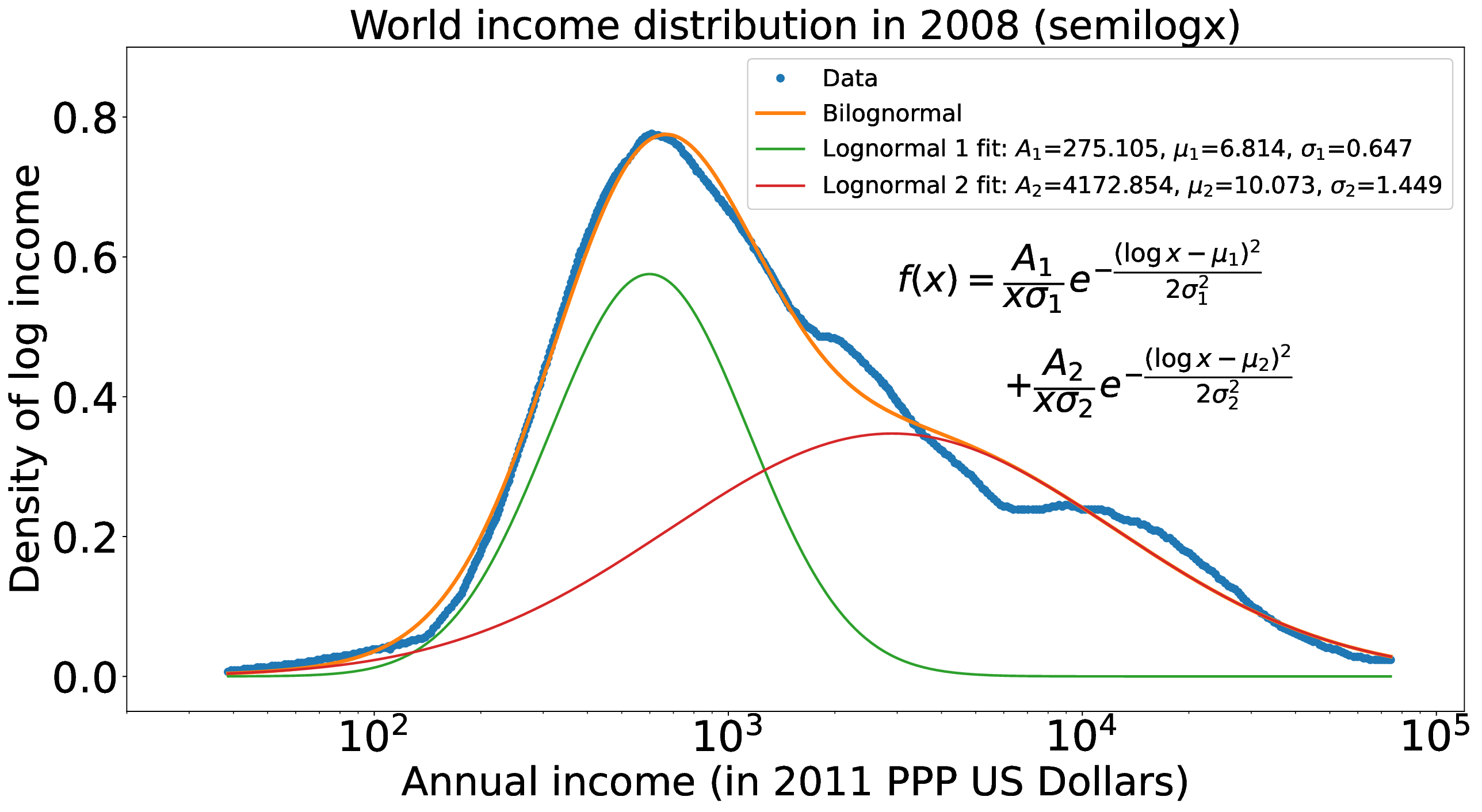}
    \caption{Same as Fig. \ref{fig:B88BL} but for 2008 $R^2=0.99129$}
    \label{fig:B08BL}
\end{figure}

\begin{figure}[ht]
    \centering
    \includegraphics[width=\linewidth]{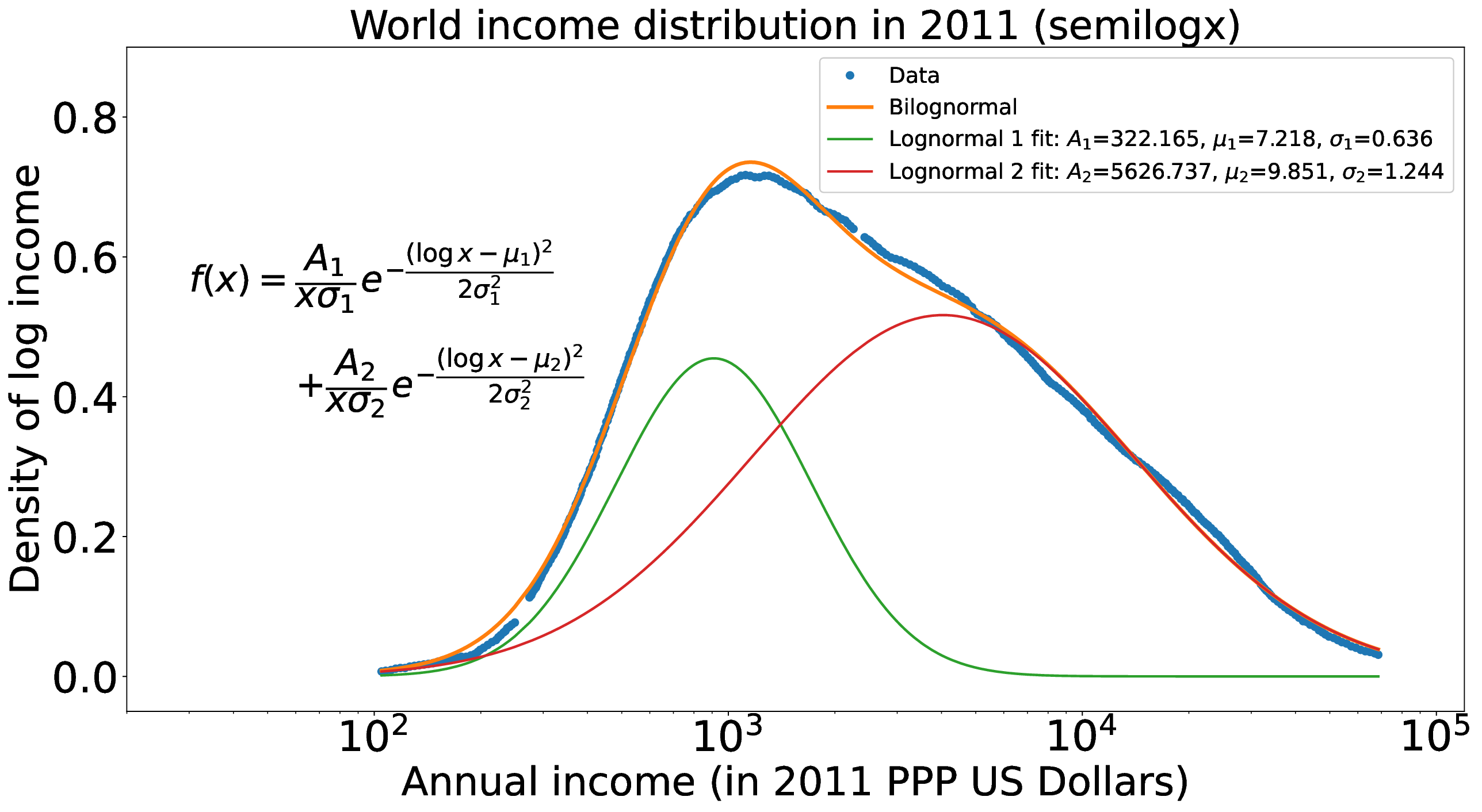}
    \caption{Same as Fig. \ref{fig:B88BL} but for 2011 $R^2=0.99729$}
    \label{fig:B11BL}
\end{figure}

\begin{figure}[ht]
    \centering
    \includegraphics[width=\linewidth]{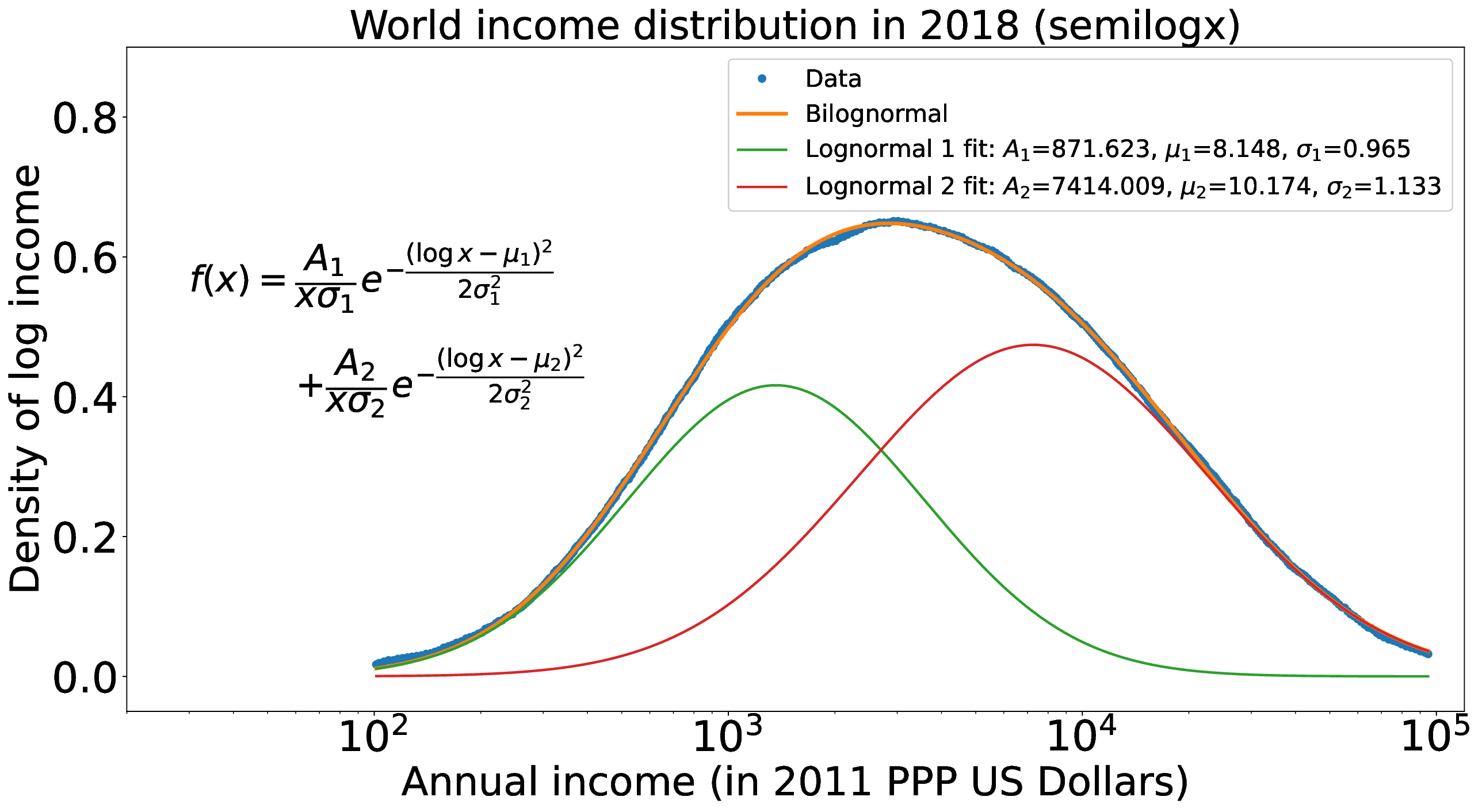}
    \caption{Same as Fig. \ref{fig:B88BL} but for 2018 $R^2=0.99973$}
    \label{fig:B18BL}
\end{figure}

\begin{figure}[ht]
    \centering
    \includegraphics[width=\linewidth]{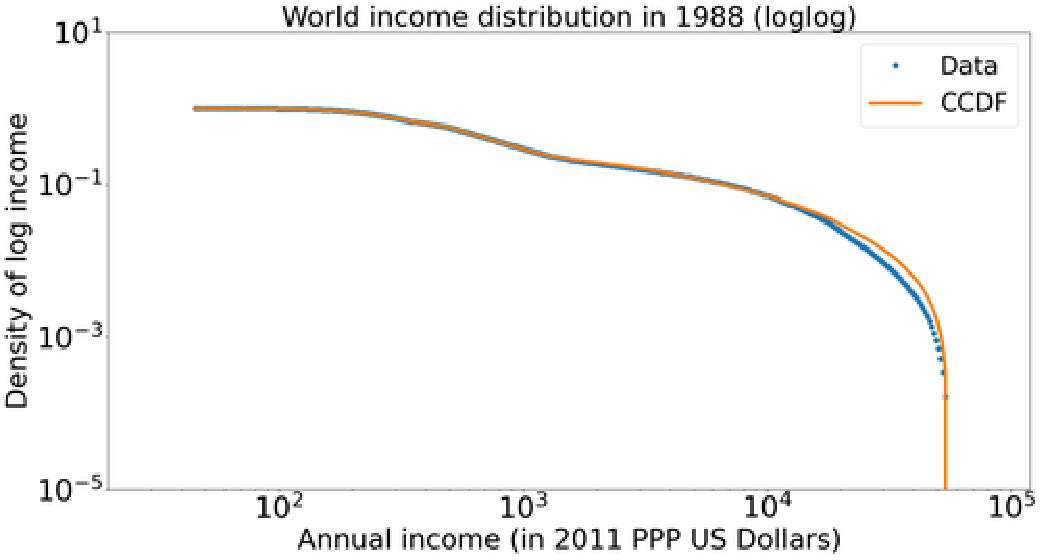}
    \caption{1988, $R^2=0.99989$}
    \label{fig:BCL88}
\end{figure}

\begin{figure}[ht]
    \centering
    \includegraphics[width=\linewidth]{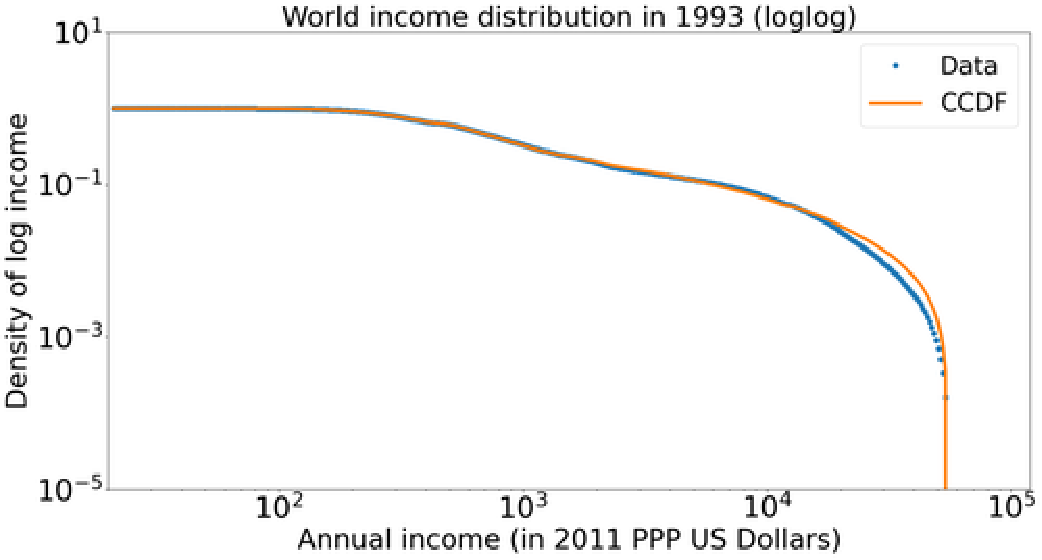}
    \caption{1993, $R^2=0.99987$}
    \label{fig:BCL93}
\end{figure}

\begin{figure}[ht]
    \centering
    \includegraphics[width=\linewidth]{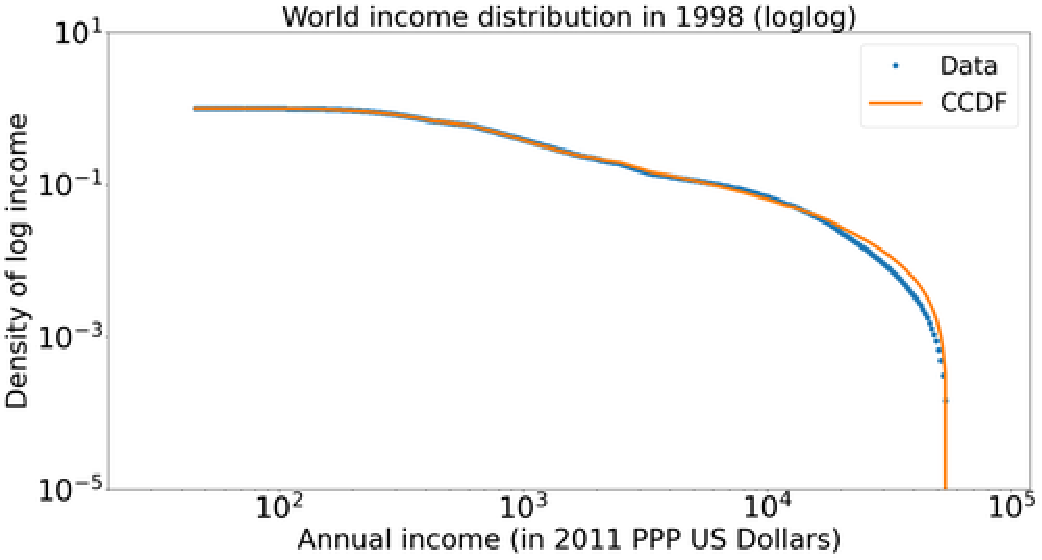}
    \caption{1998, $R^2=0.99981$}
    \label{fig:BCL98}
\end{figure}

\begin{figure}[ht]
    \centering
    \includegraphics[width=\linewidth]{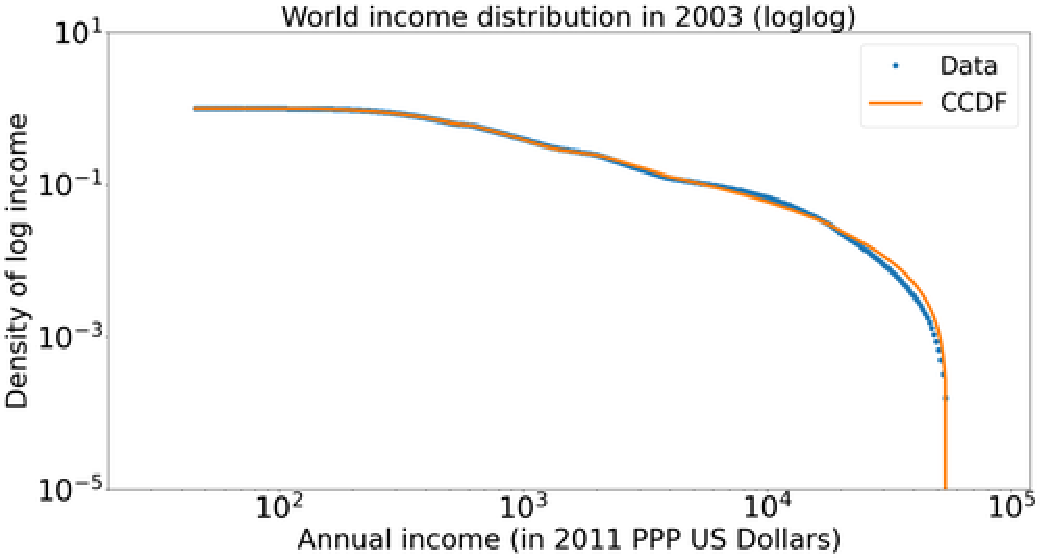}
    \caption{2003, $R^2=0.99976$}
    \label{fig:BCL03}
\end{figure}

\begin{figure}[ht]
    \centering
    \includegraphics[width=\linewidth]{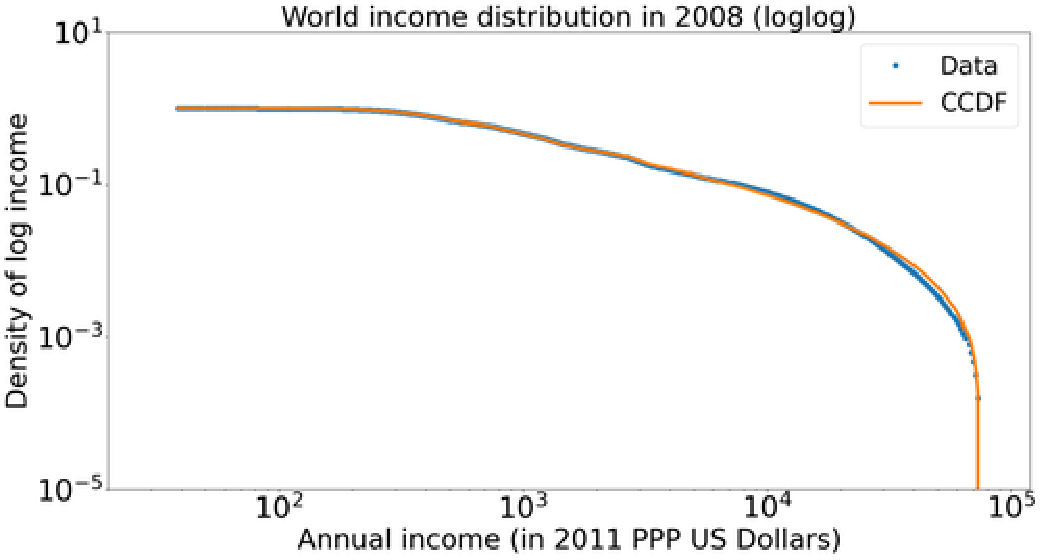}
    \caption{2008, $R^2=0.99991$}
    \label{fig:BCL08}
\end{figure}

\begin{figure}[ht]
    \centering
    \includegraphics[width=\linewidth]{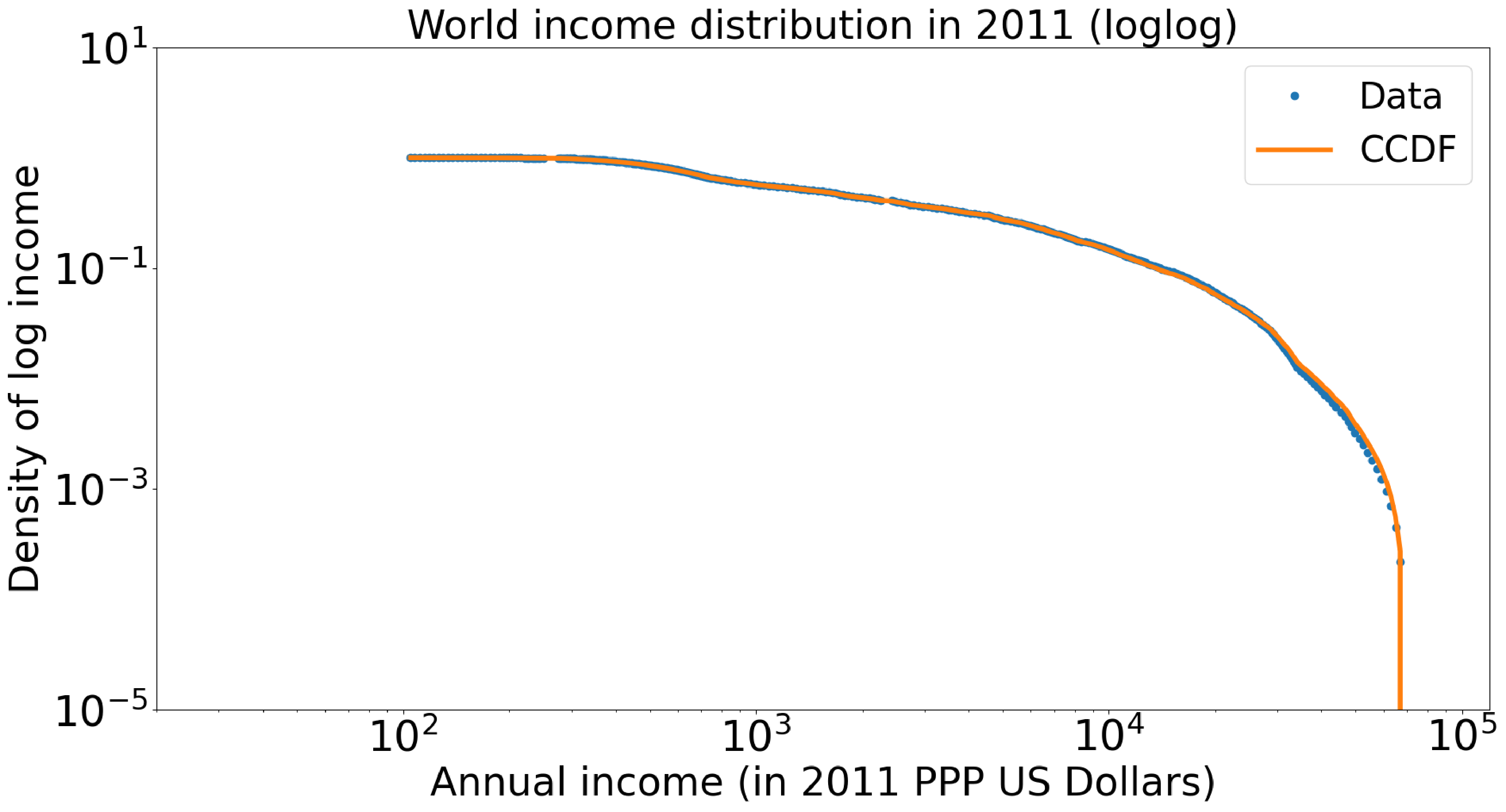}
    \caption{2011, $R^2=0.99995$}
    \label{fig:BCL11}
\end{figure}

\begin{figure}[ht]
    \centering
    \includegraphics[width=\linewidth]{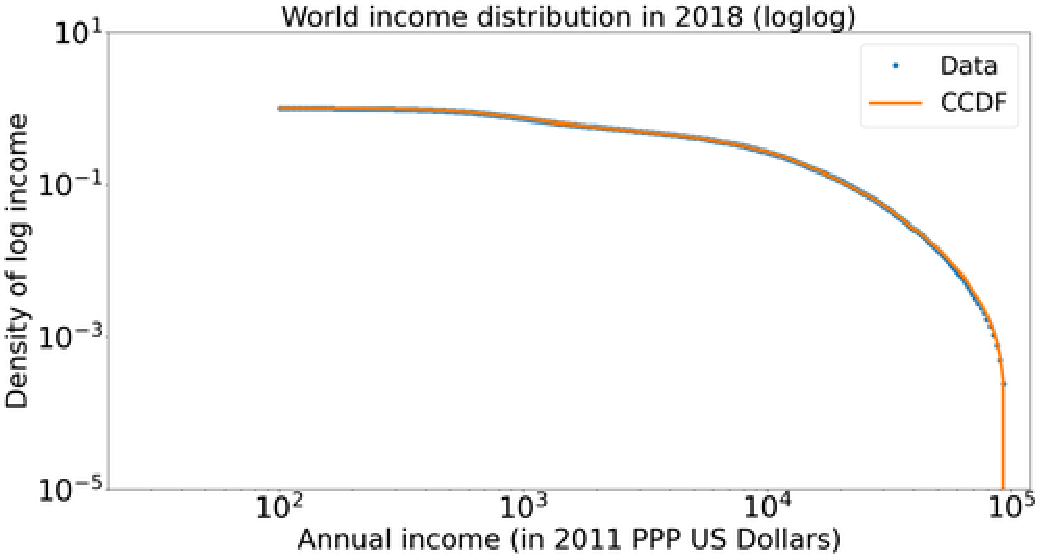}
    \caption{2018, $R^2=0.99999$}
    \label{fig:BCL18}
\end{figure}

As in the case of the gamma function, the log-normal fits shown in
Figs.\ \ref{fig:BLN88} to \ref{fig:BLN18} coincide well with all
data for the first peak. Figs.\ \ref{fig:B88BL} to \ref{fig:B18BL}
show that bi-log-normal fits are better than bi-gamma fits as shown
by the $R^2$ values shown in Table \ref{FITS}. Figs.\ \ref{fig:BCL88}
to \ref{fig:BCL18} present the respective bi-log-normal CCDF where it
is clear that this function provides a better fit at the tail of the
distribution as compared to the bi-gamma CCDF ones.


\begin{table}[p]
\footnotesize
\caption{$R^2$ values of fittings.}
\begin{tabular}{rccc}
\hline
year &  gamma  &  bi-gamma  &  bi-gamma CCDF  \\ \hline 
1988 &  0.69068  &  0.96115  &  0.99782  \\ 
1993 &  0.73948  &  0.9499  &  0.99712  \\ 
1998 &  0.72081  &  0.94064  &  0.99721  \\ 
2003 &  0.73558  &  0.93372  &  0.99711  \\ 
2008 &  0.72134  &  0.94681  &  0.99685  \\ 
2011 &  0.69562  &  0.94975  &  0.99743  \\ 
2018 &  0.85627  &  0.85627  &  0.99896  \\ \hline \\ \hline
year &  log-normal  &  bi-log-normal  &  bi-log-normal CCDF \\ \hline 
1988 &  0.77358  &  0.99383  &  0.99989   \\ 
1993 &  0.8366  &  0.98898  &  0.99987   \\ 
1998 &  0.84092  &  0.98634  &  0.99981   \\ 
2003 &  0.85684  &  0.98224  &  0.99976   \\ 
2008 &  0.86455  &  0.99129  &  0.99991   \\ 
2011 &  0.90122  &  0.99729  &  0.99995   \\ 
2018 &  0.98991  &  0.99973  &  0.99999   \\ \hline
\end{tabular}
\label{FITS}
\end{table}


\subsection{World distribution without both China and India}

\begin{figure}[ht]
    \centering
    \includegraphics[width=\linewidth]{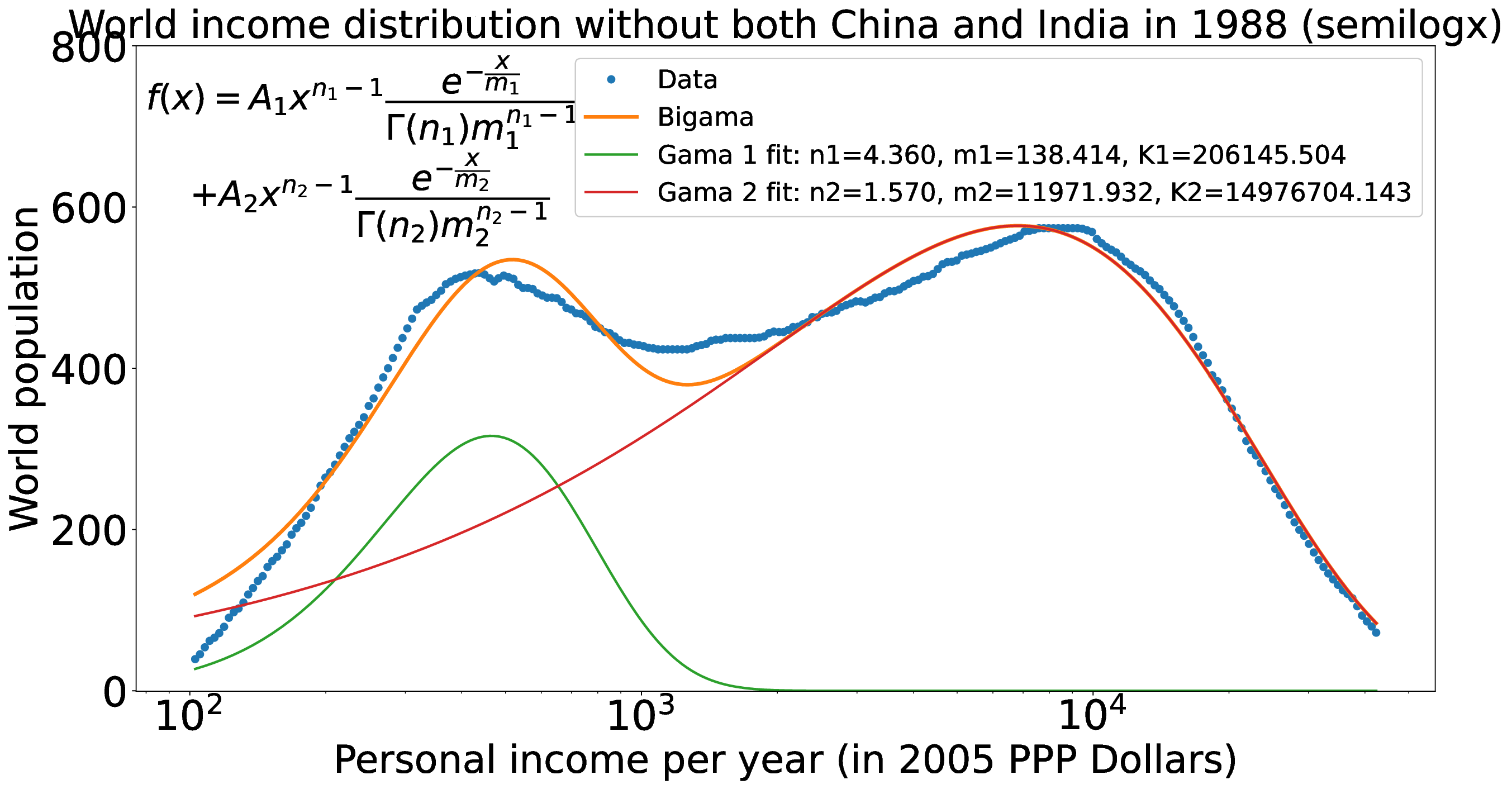}
    \caption{1988, $R^2=0.96851$}
    \label{fig:SCI88}
\end{figure}

\begin{figure}[ht]
    \centering
    \includegraphics[width=\linewidth]{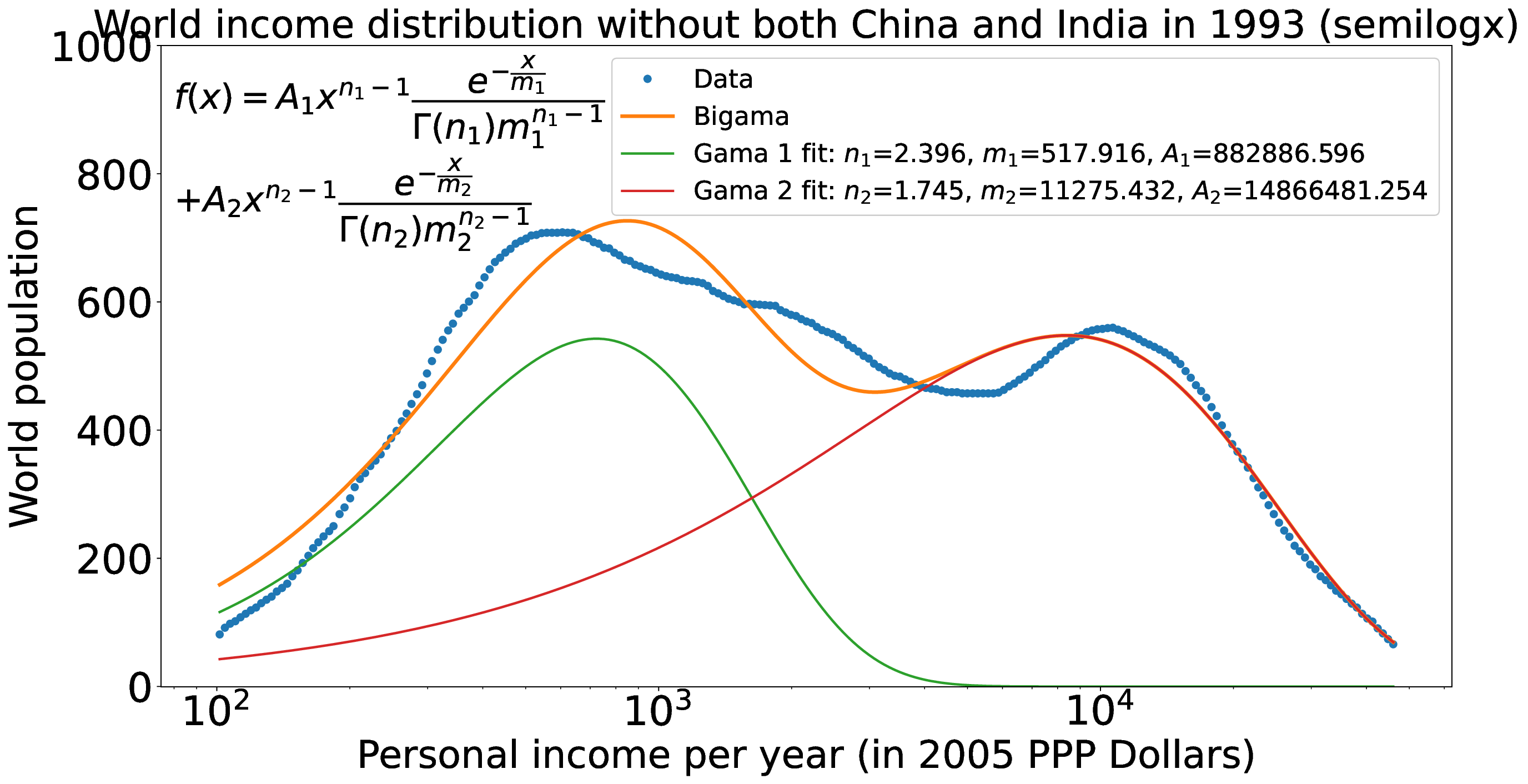}
    \caption{1993, $R^2=0.93952$}
    \label{fig:SCI93}
\end{figure}

\begin{figure}[ht]
    \centering
    \includegraphics[width=\linewidth]{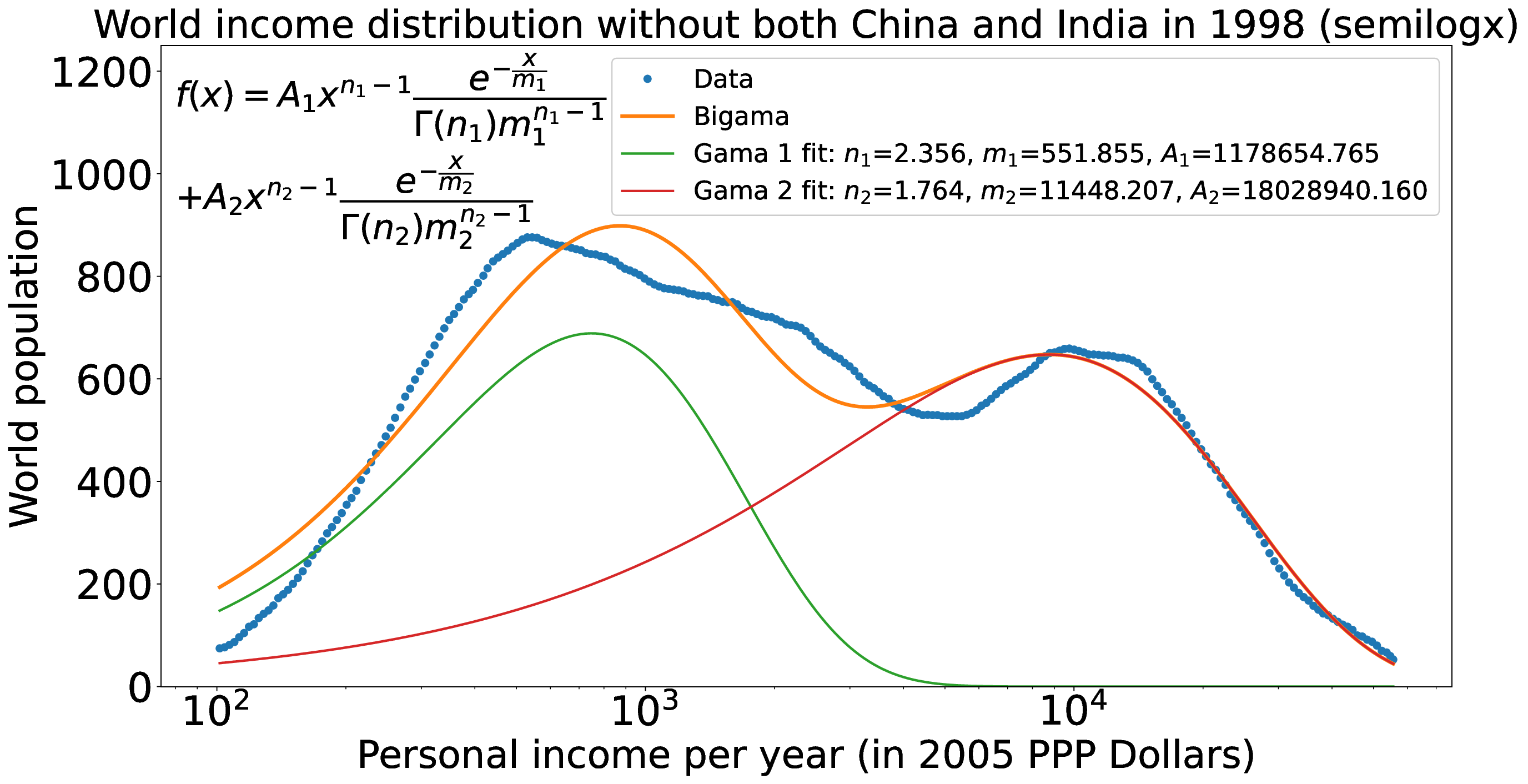}
    \caption{1998, $R^2=0.93612$}
    \label{fig:SCI98}
\end{figure}

\begin{figure}[ht]
    \centering
    \includegraphics[width=\linewidth]{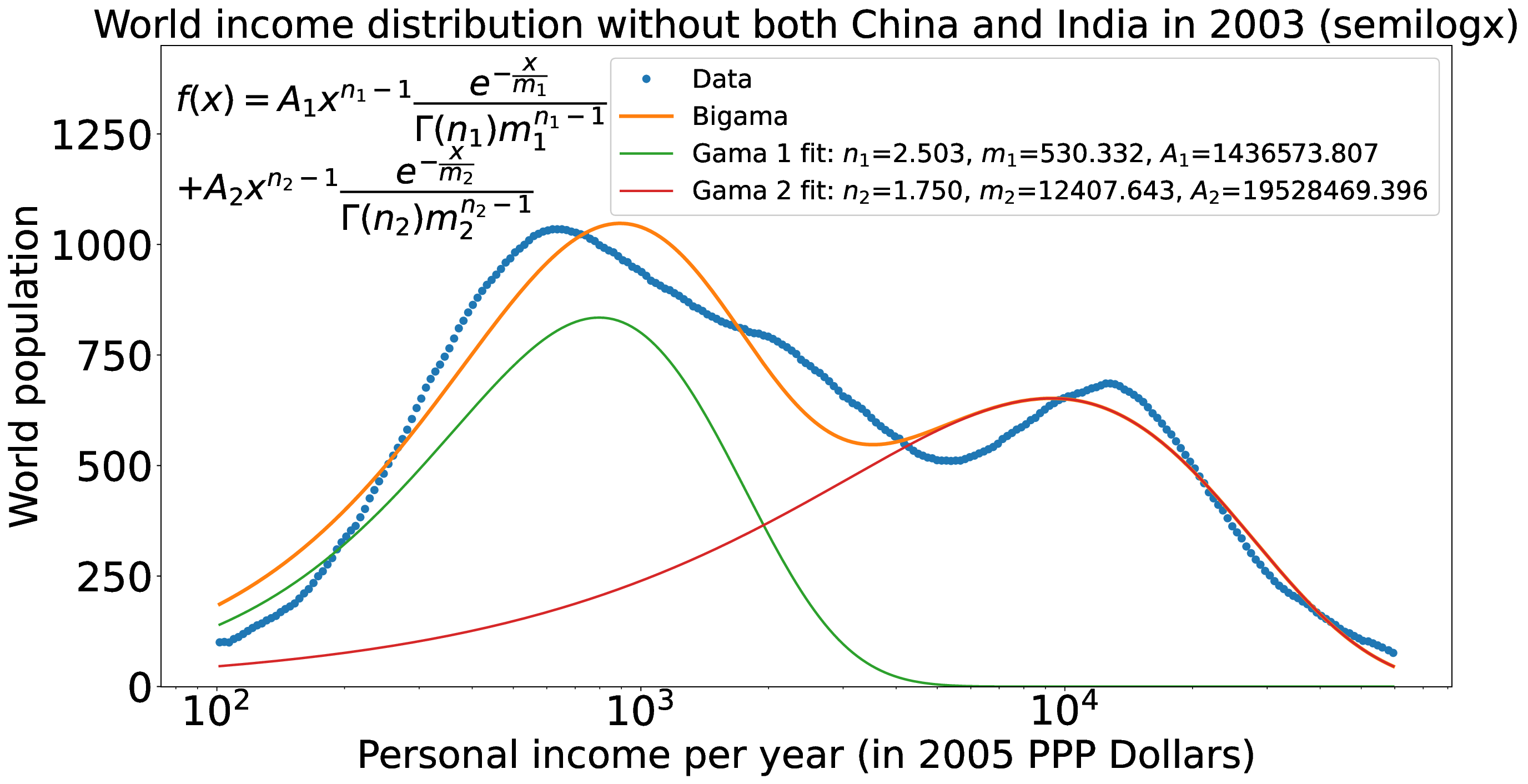}
    \caption{2003, $R^2=0.93766$}
    \label{fig:SCI03}
\end{figure}

\begin{figure}[ht]
    \centering
    \includegraphics[width=\linewidth]{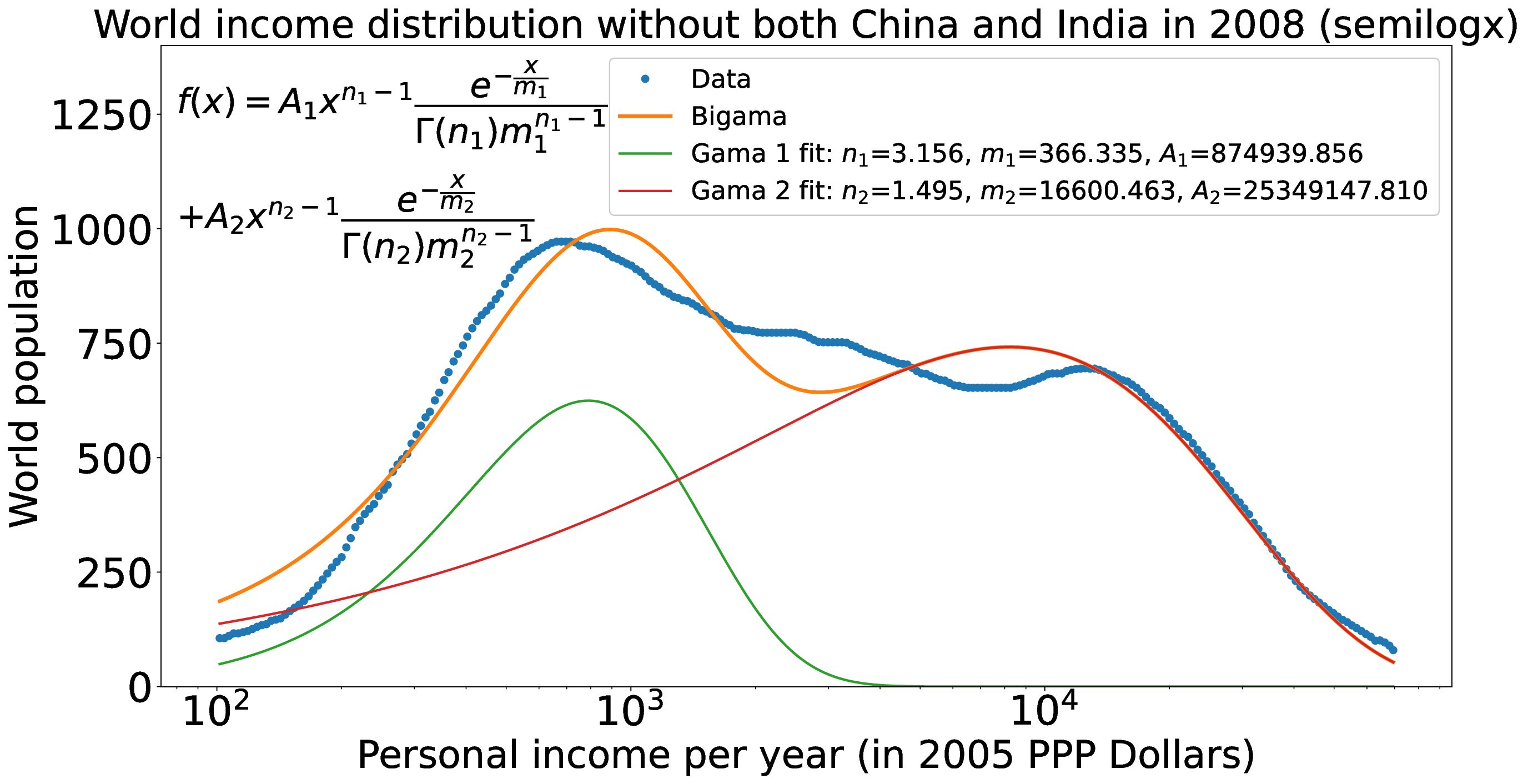}
    \caption{2008, $R^2=0.94921$}
    \label{fig:SCI08}
\end{figure}

\begin{figure}[ht]
    \centering
    \includegraphics[width=\linewidth]{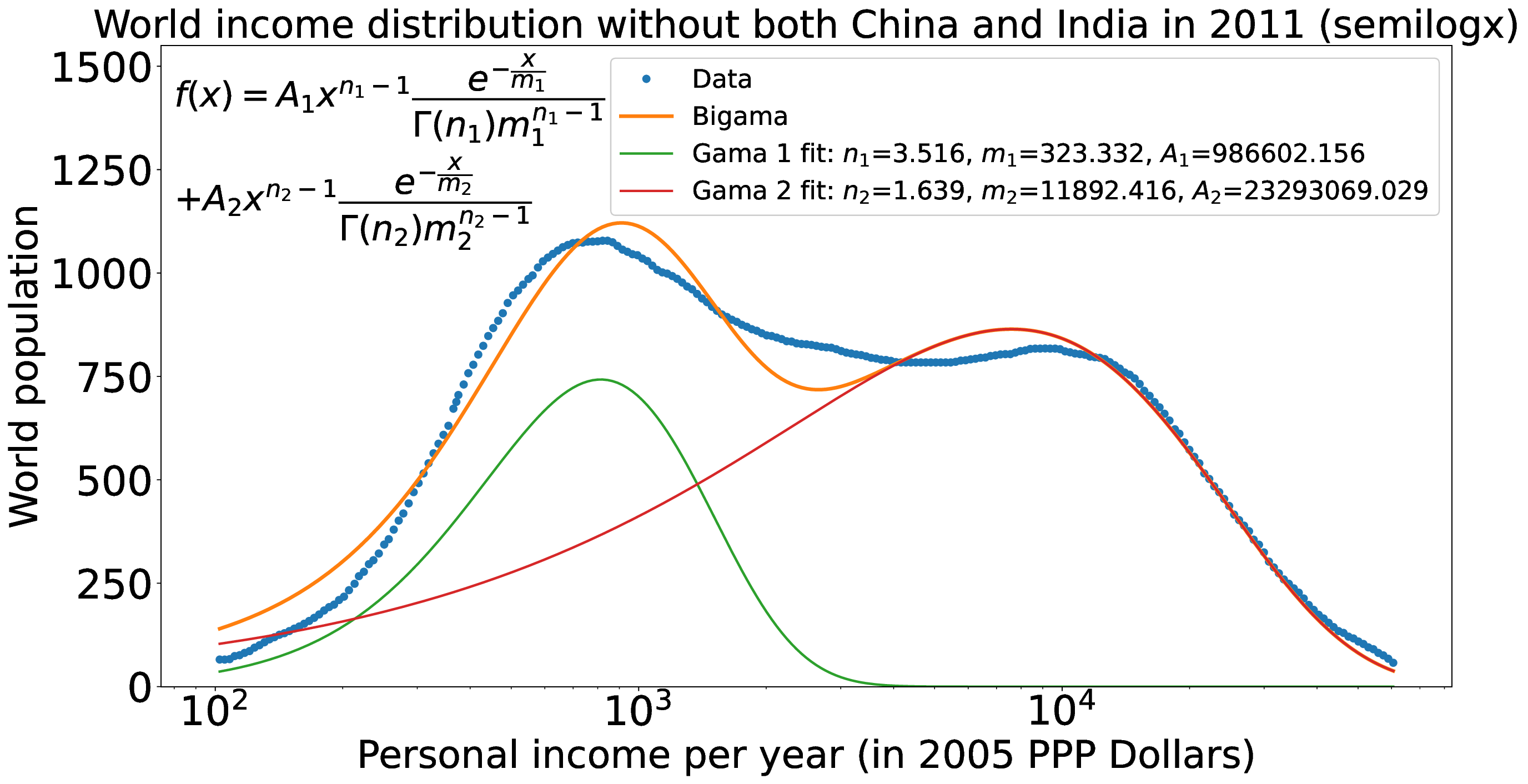}
    \caption{2011, $R^2=0.97056$}
    \label{fig:SCI11}
\end{figure}

Roser \cite{Roser2023-et} provided the income distributions of China
and India along time, and this allowed us to conveniently subtract
their contribution to the global distribution after realizing these
countries play a fundamental role in shaping global distribution.
Figs.\ \ref{fig:SCI88} to \ref{fig:SCI11} present these results where
the Y-axis is the PDF of the fitted functions.

If we envision a scenario without the presence of these two significant
demographic and economic players, a pronounced decline in the poor and
middle-class values emerge, creating a noticeable ``valley'' in the
graph. So, it seems that China and India, with their vast populations
and expanding economies, act as a bridge that fills this valley, thereby
generating a more uniform and comprehensive data distribution. 

A comparative analysis shows that while earlier works, such as those by
Milanovic, focused on single-peak representations of income distribution,
they often failed to capture the complexities introduced by bimodal patterns,
especially in global datasets. Similarly, many works in econophysics
highlighted exponential and power-law distributions for national economies
but to our knowledge, but did not address the multimodal characteristics
observed in global contexts. 

Therefore, our bimodal adjustments provide a new way of fitting the world
income distribution along the years (qualified with R values) showing the
usefulness of combining multiple distributions to model diverse economic
systems effectively. Insights gained from our models can provide elegant
interpretations of broader economic trends that can be improved by employing
distributions with more parameters.

\section{Conclusions}
This paper analyzed the evolution of global income distributions over
several decades using empirical datasets and fits to standard statistical
functions: gamma, log-normal, and their bimodal combinations. The graphical
analysis of the \textit{probability density function} (PDF) and the
\textit{complementary cumulative density function} (CCDF) revealed clear
patterns of inequality and the presence of multimodality in global income.

A key result is that single-function fits (gamma or log-normal) yield
reasonably good approximations up to \$60k (2011 PPP USD), but the use
of bimodal combinations significantly improves fit quality across all years
studied. This suggests the global income distribution is better described
as a composition of at least two subpopulations, reflecting different
economic realities. The goodness-of-fit, measured by $R^2$, supports this
conclusion.

More recently, although there were improvements in collecting datasets
for income and wealth in many countries, data for the very poor and the
very rich households and individuals are still, in general, quite
unreliable. Due to the above mentioned upper limit data cutoff most
people from poor countries, with low per capita GDP, were included, but
in places like Monaco, Qatar and others with relatively large income per
capita, the datasets do not include large percentages of their respective
populations. The absence of significant world high-income datasets
explains why the Pareto power-law is very poorly presented, or
non-existent, in various upper segments of the income distributions
showed here. Therefore, world income distributions with higher upper
limits should have exhibited clear power law behaviors for higher income
values in addition to, possibly, second or third similar power-law
behaviors in subsets of increasing incomes. For instance, the income
and wealth data for billionaires as given by Forbes magazine
\cite{noauthor_undated-wn} every year represents a very small subset of
humans who are very important in studies of economic income and wealth
inequalities related to the so-called, and much talked about, 99\% vs.\
1\% economic disparity in the whole world.

A particularly novel finding is the structural role of China and India
in shaping the global income distribution. Between 1988 and 2008, the
global distribution transitioned from a bimodal to a more unimodal form,
largely due to income growth in these populous nations. When these two
countries are excluded, a valley between low- and middle-income peaks
re-emerges, underscoring their bridging function in the global economy.

These results underscore the value of combining multiple distributions to
better capture global economic heterogeneity. Moreover, they open the
door to richer interpretations of macroeconomic trends and transitions
in global inequality.

Importantly, such modeling approaches can be applied to study the effects
of shocks or policy interventions. For example, by fitting similar
distributions to post-pandemic datasets, one could quantify the impact
of COVID-19 on income polarization or middle-class shrinkage in different
regions, an issue already under debate in recent literature
\cite{neelima2022post,UNDP2022}.

This work provides a simple and elegant comprehensive approach to the
world income distributions fitted to commonly used functions in econophysics.

The implications of these findings pave the way for more refined studies
and expanded datasets. The following section outlines potential directions
to deepen our understanding of income dynamics and possible policy implications.

\section{Future work}

We are particularly interested in extending the present analysis employing
very recent data from the LIS database \cite{LIS}, which contain household
-and person- level data on labor income, capital income, pensions, public
social benefits and private transfers, as well as taxes and contributions,
demography, employment, and expenditures. So, this database will provide
enough empirical results to perform many important  comparative
interdisciplinary analysis.

\section*{Acknowledgments}
J.D.A.I.G.\ and M.d.C.M.\ acknowledge the partial financial support
provided by DGAPA-UNAM, Mexico, grant No.\ IN106225. M.B.R.\ received
partial financial support from FAPERJ -- \textit{Rio de Janeiro State
Research Funding Agency}, grant number E-26/210.552/2024.

\bibliographystyle{elsarticle-num}
\bibliography{econophys}

\def\cprime{$'$}
\begin{thebibliography}{10}
\expandafter\ifx\csname url\endcsname\relax
  \def\url#1{\texttt{#1}}\fi
\expandafter\ifx\csname urlprefix\endcsname\relax\def\urlprefix{URL }\fi
\expandafter\ifx\csname href\endcsname\relax
  \def\href#1#2{#2} \def\path#1{#1}\fi

\bibitem{oxfam1}
A.~Th\'{e}riault, Richest 1\% bag nearly twice as much wealth as the rest of
  the world put together over the past two years, Press release, Oxfam
  International,
  \url{https://www.oxfam.org/en/press-releases/richest-1-bag-nearly-twice-much-wealth-rest-world-put-together-over-past-two-years},
  (accessed 17 June 2024) (2023).

\bibitem{oxfam2}
M.-B. Christensen, C.~Hallum, A.~Maitland, Q.~Parrinello, C.~Putaturo, D.~Abed,
  C.~Brown, A.~Kamande, M.~Lawson, S.~Ruiz,
  \href{www.oxfam.org/en/research/survival-richest}{Survival of the richest},
  Policy paper (last accessed 29 {A}ugust 2024), Oxfam International (2023).
\newline\urlprefix\url{www.oxfam.org/en/research/survival-richest}

\bibitem{piketty}
T.~Piketty, Capital in the Twenty-First Century, Harvard University Press,
  Cambridge, Massachusetts, 2014, translated by Arthur Goldhammer from the
  author's original in French \textit{Le capital au XXIe si\'ecle}.

\bibitem{incomedistro}
B.~K. Chakrabarti, A.~Chakraborti, S.~T. Chakravarty, A.~Chatterjee,
  Econophysics of Income and Wealth Distributions, Cambridge University Press,
  Cambridge, 2013.

\bibitem{ribeiro2020income}
M.~B. Ribeiro, Income Distribution Dynamics of Economic Systems: An
  Econophysical Approach, Cambridge University Press, Cambridge, 2020.

\bibitem{elephantpaper2}
C.~Lakner, B.~Milanovic, Global {I}ncome {D}istribution: From the {F}all of the
  {B}erlin {W}all to the {G}reat {R}ecession, The World Bank Economic Review
  30~(2) (2015) 203--232.
\newblock \href {https://doi.org/10.1093/wber/lhv039}
  {\path{doi:10.1093/wber/lhv039}}.

\bibitem{milanovic2002true}
B.~Milanovic, True world income distribution, 1988 and 1993: First calculation
  based on household surveys alone, The economic journal 112~(476) (2002)
  51--92.

\bibitem{milanovic2016greatest}
B.~Milanovic, The greatest reshuffle of individual incomes since the industrial
  revolution, VoxEU July 1 (2016) 2016.

\bibitem{anand2015global}
S.~Anand, P.~Segal, The global distribution of income, in: Handbook of income
  distribution, Vol.~2, Elsevier, 2015, pp. 937--979.

\bibitem{bourguignon2002size}
F.~Bourguignon, C.~Morrisson, The size distribution of income among world
  citizens, 1820-1990, American Economic Review 92~(4) (2002) 727--44.

\bibitem{berry1983changes}
A.~Berry, F.~Bourguignon, C.~Morrison, Changes in the world distribution of
  income between 1950 and 1977, The Economic Journal 93~(370) (1983) 331--350.

\bibitem{milanovic2022after}
B.~Milanovic, After the financial crisis: the evolution of the global income
  distribution between 2008 and 2013, Review of Income and Wealth 68~(1) (2022)
  43--73.

\bibitem{milanovic2006global}
B.~Milanovic, Global income inequality: What it is and why it matters (2006).

\bibitem{milanovic2011worlds}
B.~Milanovic, Worlds apart: Measuring international and global inequality,
  Princeton University Press, 2011.

\bibitem{milanovic2012global}
B.~Milanovic, Global inequality recalculated and updated: the effect of new ppp
  estimates on global inequality and 2005 estimates, The Journal of Economic
  Inequality 10 (2012) 1--18.

\bibitem{milanovic2016income}
B.~Milanovic, Income inequality is cyclical, Nature 537~(7621) (2016) 479--482.

\bibitem{milanovic2023great}
B.~Milanovic, The great convergence: Global equality and its discontents,
  Foreign Aff. 102 (2023) 78.

\bibitem{anand2008we}
S.~Anand, P.~Segal, What do we know about global income inequality?, Journal of
  Economic Literature 46~(1) (2008) 57--94.

\bibitem{anand2010debates}
S.~Anand, P.~Segal, J.~E. Stiglitz, Debates on the measurement of global
  poverty, Oxford University Press, 2010.

\bibitem{anand2017global}
S.~Anand, P.~Segal, Who are the global top 1\%?, World Development 95 (2017)
  111--126.

\bibitem{segal2022inequality}
P.~Segal, Inequality interactions: The dynamics of multidimensional
  inequalities, Development and Change 53~(5) (2022) 941--961.

\bibitem{atkinson97}
A.~B. Atkinson, Bringing income distribution in from the cold, The Economic
  Journal 107 (1997) 297--321.

\bibitem{pareto1964cours}
V.~Pareto, Cours d'\'{E}conomie Politique, Vol.~2, F.\ Rouge, Lausanne, 1897.

\bibitem{gibrat1931inegalits}
R.~Gibrat, Les in{\'e}galits {\'e}conomiques, Sirey (1931).

\bibitem{yakovenko2005two}
V.~M. Yakovenko, A.~C. Silva, Two-class structure of income distribution in the
  usa: Exponential bulk and power-law tail, in: Econophysics of Wealth
  Distributions: Econophys-Kolkata I, Springer, 2005, pp. 15--23.

\bibitem{silva2004temporal}
A.~C. Silva, V.~M. Yakovenko, Temporal evolution of the “thermal” and
  “superthermal” income classes in the usa during 1983--2001, Europhysics
  Letters 69~(2) (2004) 304.

\bibitem{soriano2017non}
P.~Soriano-Hern{\'a}ndez, M.~del Castillo-Mussot,
  O.~C{\'o}rdoba-Rodr{\'\i}guez, R.~Mansilla-Corona, Non-stationary individual
  and household income of poor, rich and middle classes in mexico, Physica A:
  Statistical Mechanics and its Applications 465 (2017) 403--413.

\bibitem{jagielski2013modelling}
M.~Jagielski, R.~Kutner, Modelling of income distribution in the european union
  with the fokker--planck equation, Physica A: Statistical Mechanics and its
  Applications 392~(9) (2013) 2130--2138.

\bibitem{tao2019exponential}
Y.~Tao, X.~Wu, T.~Zhou, W.~Yan, Y.~Huang, H.~Yu, B.~Mondal, V.~M. Yakovenko,
  Exponential structure of income inequality: evidence from 67 countries,
  Journal of Economic Interaction and Coordination 14 (2019) 345--376.

\bibitem{joseph2022}
B.~Joseph, B.~K. Chakrabarti, Variation of {G}ini and {K}olkata indices with
  saving propensity in the kinetic exchange model of wealth distribution: An
  analytical study, Physica A 594 (2022) 127051.
\newblock \href {https://doi.org/10.1016/j.physa.2022.127051}
  {\path{doi:10.1016/j.physa.2022.127051}}.

\bibitem{survive}
D.~F. Moore, Applied Survival Analysis Using R, Springer, 2016.

\bibitem{reli}
M.~Mohammad~Modarres, M.~Kaminskiy, V.~Krivtsov, Reliability Engineering and
  Risk Analysis: A Practical Guide, Marcel Dekker, Inc., 1999.

\bibitem{kleiber2003statistical}
C.~Kleiber, S.~Kotz, Statistical size distributions in economics and actuarial
  sciences, John Wiley \& Sons, 2003.

\bibitem{ferrero2004statistical}
J.~C. Ferrero, The statistical distribution of money and the rate of money
  transference, Physica A: Statistical Mechanics and its Applications 341
  (2004) 575--585.

\bibitem{ferrero2005monomodal}
J.~C. Ferrero, The monomodal, polymodal, equilibrium and nonequilibrium
  distribution of money, in: Econophysics of Wealth Distributions:
  Econophys-Kolkata I, Springer, 2005, pp. 159--167.

\bibitem{lakner2015global}
C.~Lakner, B.~Milanovic, Global income distribution from the fall of the berlin
  wall to the great recession, Revista de Econom{\'\i}a Institucional 17~(32)
  (2015) 71--128.

\bibitem{Roser2023-et}
M.~Roser, The history of global economic inequality, 'The history of global
  economic inequality'. Published online at OurWorldInData.org. Retrieved from:
  'https://ourworldindata.org/the-history-of-global-economic-inequality'
  [Online Resource] (2017).

\bibitem{noauthor_undated-wn}
The world's real-time billionaires, \\\url{https://www.forbes.com/real-time-}
  \\\url{billionaires/#321e71413d78}, (accessed 5 September 2023) (2024).

\bibitem{neelima2022post}
K.~Neelima, S.~Chennapalli, Post-pandemic global inequalities: Causes and
  measures, in: Emerging Trends and Insights on Economic Inequality in the Wake
  of Global Crises, IGI Global, 2022, pp. 40--55.

\bibitem{UNDP2022}
U.~N.~D. Programme,
  \href{https://hdr.undp.org/content/human-development-report-2021-22}{Uncertain
  times, unsettled lives: Shaping our future in a transforming world} (2022).
\newline\urlprefix\url{https://hdr.undp.org/content/human-development-report-2021-22}

\bibitem{LIS}
\href{https://www.lisdatacenter.org/our-data/lis-database/}{The {L}uxembourg
  {I}ncome {S}tudy {D}atabase ({LIS})}.
\newline\urlprefix\url{https://www.lisdatacenter.org/our-data/lis-database/}

\end{thebibliography}
\end{document}